\documentclass{jfm}

\usepackage{etex}
\usepackage{lmodern}
\usepackage{gensymb}
\usepackage[dvips]{graphicx}
\usepackage{url}
\usepackage{setspace}
\usepackage{natbib}
\usepackage[percent]{overpic}
\usepackage{graphics}
\usepackage{amssymb}
\usepackage{amsmath}
\usepackage{mathrsfs}
\usepackage{mathtools}
\usepackage{cancel}
\usepackage{soul}
\usepackage{subfig}
\usepackage{float}
\usepackage{textcomp}
\usepackage{listings}
\usepackage{placeins}
\usepackage{color}
\usepackage{epsfig}
\usepackage{bm}
\usepackage{lscape}
\usepackage[usenames,dvipsnames,table]{xcolor}
\usepackage{tikz}
\usetikzlibrary{positioning,arrows}
\usetikzlibrary{decorations.pathreplacing}
\usepackage{pstricks}
\usepackage[toc,page]{appendix}
\usepackage{upgreek}
\usepackage{caption}
\usepackage[mathlines]{lineno}
\usepackage{hyperref}% http://ctan.org/pkg/hyperref
\usepackage{soul}
\allowdisplaybreaks
\usepackage{lineno}
\usepackage{paralist}
\usepackage{makecell}

\ifCUPmtlplainloaded \else
  \checkfont{eurm10}
  \iffontfound
    \IfFileExists{upmath.sty}
      {\typeout{^^JFound AMS Euler Roman fonts on the system,
                   using the 'upmath' package.^^J}%
       \usepackage{upmath}}
      {\typeout{^^JFound AMS Euler Roman fonts on the system, but you
                   dont seem to have the}%
       \typeout{'upmath' package installed. JFM.cls can take advantage
                 of these fonts,^^Jif you use 'upmath' package.^^J}%
      }
  \else
  \fi
\fi

\ifCUPmtlplainloaded \else
  \checkfont{msam10}
  \iffontfound
    \IfFileExists{amssymb.sty}
      {\typeout{^^JFound AMS Symbol fonts on the system, using the
                'amssymb' package.^^J}%
       \usepackage{amssymb}%
       \let\le=\leqslant  
       \let\ge=\geqslant  \let\geq=\geqslant
      }{}
  \fi
\fi

\ifCUPmtlplainloaded \else
  \IfFileExists{amsbsy.sty}
    {\typeout{^^JFound the 'amsbsy' package on the system, using it.^^J}%
     \usepackage{amsbsy}}
    {}
\fi

\newcommand\p{\ensuremath{\partial}}

\newcommand{\beq}{\begin{equation}}
\newcommand{\eeq}{\end{equation}}
\newcommand{\dissip}{\ensuremath{\mathcal{D}}} % dissipation

\title[Stabilisation and drag reduction of pipe flows]{
Stabilisation and drag reduction of pipe flows by flattening the base profile}

\author[E. Marensi, A. P. Willis and R. R. Kerswell]%
{Elena Marensi$^1$\thanks{Email address for correspondence: e.marensi@sheffield.ac.uk}, Ashley P. Willis$^1$, Rich R. Kerswell$^2$}

\affiliation{$^1$School of Mathematics and Statistics, University of Sheffield, Sheffield S3 7RH, UK\\
$^2$Centre for Mathematical Sciences, University of Cambridge, Cambridge CB3 0WA, UK
}

\pubyear{}
\volume{}
\pagerange{}
\begin{document}

\maketitle

\begin{abstract}
Recent experimental observations (K{\"u}hnen {\it{et al.}}, {\it{Nat. Phys.}}, 2018) have shown that flattening a turbulent streamwise velocity profile in pipe flow destabilises the turbulence so that the flow relaminarises. We show that a similar phenomenon exists for laminar pipe flow profiles in the sense that the nonlinear stability of the laminar state is enhanced as the profile becomes more flattened. Significant drag reduction is also observed for the turbulent flow when triggered by sufficiently large disturbances. The flattening of the laminar base profile is produced by an artificial localised body force designed to mimic an obstacle used in the experiments of K{\"u}hnen {\it{et al.}} ({\it{Flow Turbul. Combust.}}, 2018) and the nonlinear stability measured by the size of the energy of the initial perturbations needed to trigger transition. In order to make the latter computation more efficient, we examine how indicative the minimal seed $-$ the disturbance of smallest energy for transition $-$ is in measuring transition thresholds. We first show that the minimal seed is relatively robust to base profile changes and spectral filtering. We then compare the (unforced) transition behaviour of the minimal seed with several forms of randomised initial conditions in the range of Reynolds numbers $Re=2400$ to $10000$ and find that the energy of the minimal seed after the Orr and oblique phases of its evolution is close to that of a localised random disturbance. In this sense, the minimal seed at the end of the oblique phase can be regarded as a good proxy for typical disturbances (here taken to be the localised random ones) and is thus used as initial condition in the simulations with the body force. The enhanced nonlinear stability and drag reduction predicted in the present study are an encouraging first step in modelling the experiments of K{\"u}hnen {\it{et al.}} and should motivate future developments to fully exploit the benefits of this promising direction for flow control.
\end{abstract}

%%%%%%%%%%%%%%%%%%%%%%%%%%%%%%%%%%
\section{Introduction}
\label{intro}
It is widely established that turbulent wall flows exert a much higher friction drag than laminar flows. Since the flow regime in oil and gas pipelines is generally turbulent, larger pumping forces are needed, as compared to the laminar case, to maintain the desired flow rates, with consequent increase in energy consumption and carbon emissions. A great deal of research effort is thus directed towards the design of efficient control strategies to either reduce the turbulent drag or to delay the onset of turbulence. Transition to turbulence in pipe flows is a fully nonlinear problem because the laminar state is linearly stable to any infinitesimal disturbance. Therefore, if one wishes to control or delay transition, it is paramount to understand which kind of small (but finite-amplitude) disturbances are most effective in initiating the transition process. A useful tool that has recently been employed to tackle this challenge is the so-called minimal seed, i.e. the disturbance of lowest energy capable to trigger transition. However, the question of how representative the minimal seed is of typical ambient disturbances, remains unanswered. To address this issue, we compare the transition behaviour of the minimal seed with that of different random initial disturbances in the range of Reynolds numbers $Re=2400$ to $10000$. We find that the energy of the minimal seed after the initial nonlinear unpacking phase is quite close to that of a localised random disturbance. Suitable initial conditions are thus generated to investigate the stabilising effect of a simple model for the presence of a baffle in the core of the flow.

Before discussing the formulation and results, (refer to \S \ref{sec:formulation} and \S \ref{sec:results}, respectively) we provide a short review of the problem of transition in pipe flows and the different control strategies used to suppress it. 

\subsection{Transition in pipe flows and calculation of the minimal seed}
\label{intro:transition}
The enigma of how laminar flow through a pipe undergoes the transition to turbulence has been intriguing and challenging scientists for over a century, since the pioneering experiments of O. Reynolds in 1883 \citep{R1883}. Despite many pieces of the puzzle being brought together in the past years \citep[refer, for example, to][for comprehensive reviews]{Kerswell05,ESHW07,WPKM08,Mul10}, a full understanding of the problem still eludes us.

All theoretical and numerical evidence indicates that the laminar state is linearly stable to any infinitesimal disturbance, although a rigorous proof is still lacking.
In the absence of a linear instability of the laminar state from which a sequence of bifurcations may be initiated, transition can only be triggered by finite-amplitude background disturbances. For $Re > 3000 $ the observed transition process is abrupt and catastrophic and it rapidly results in a complex and highly disordered state \citep{darbyshire-mullin-1995}. 
At transitional Reynolds numbers in the range $1800<Re<3000$, instead, turbulence first appears in localised patches of disordered motion, known as puffs, which coexist with the laminar flow \citep{WygCha73,AMdABH11}. Depending on the level of background noise in the experiment, the flow rate at which transition occurs can be varied by more than an order of magnitude. This fact already puzzled Reynolds in 1883 who, in one set of experiments, found a transitional Reynolds number $Re_c\approx2000$, while in another set of experiments with minimised level of background disturbance, found $Re_c\approx 13000$. This value was pushed to $10^5$ by Pfenninger \citep{Pfenninger61} with a very tightly controlled environment of his experiments. Reynolds' lower critical value has been confirmed in other experiments \citep[e.g.][]{WygCha73,darbyshire-mullin-1995,AMdABH11} with current estimates in the range $1760-2300$.

At lower Reynolds numbers ($Re\approx 2000$), the critical Reynolds number is somewhat dependent on the definition of `transition', but at larger $Re$ ($Re\gtrapprox 3000$) where the transition is clear, it is widely recognised that the influence of background disturbances becomes of great concern. The critical amplitude $A_c$ for the onset of turbulence is expected to decrease with Reynolds number and its behaviour can be characterised by $A_c\sim Re^{-\gamma}$, $\gamma>0$. A key question is thus: what is the exponent $\gamma$ and, more importantly, can this value be predicted theoretically? \citet{trefethen-etal-2000} proposed a renormalisation of the amplitude by the average velocity in order to cast different experimental results in terms of a single definition of $A_c$, suggesting lower and upper bounds for $\gamma \in [6/5, 9/5]$. The experiments carried out by \citet{peixinho-mullin-2007} provided a critical exponent $\gamma \in [1.3, 1.5]$ when the flow was perturbed using push-pull disturbances and $\gamma=1$ when the flow was perturbed by small impulsive jets. The latter scaling had previously been found in the experiments of \citet{HJM03} and was later confirmed numerically by \citet{mellibovsky-meseguer-2009} in their `impulsive scenario' with the flow being perturbed by a local impulsive forcing.
In the `autonomous scenario', instead, where the flow was perturbed by an initial array of streamwise vortices with random noise superimposed on it, \citet{mellibovsky-meseguer-2009} obtained critical exponents $\gamma \in [1, 1.5]$ much closer to those of \citet{peixinho-mullin-2007} for the push-pull disturbances.

\begin{figure}
  \centering
	\includegraphics[width=0.5 \textwidth]{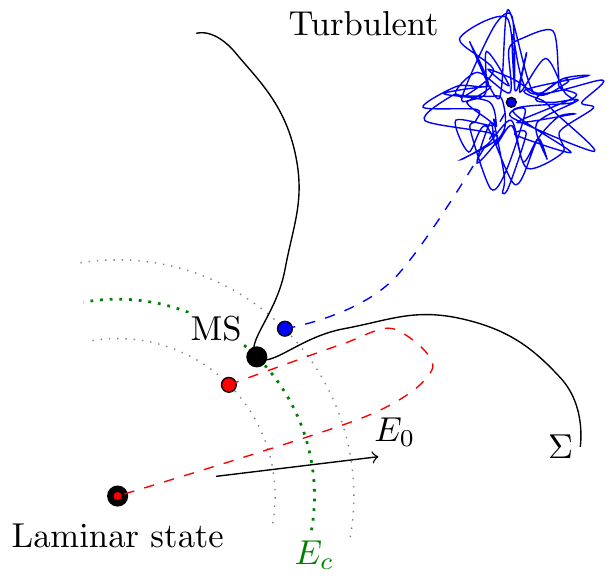}
  \caption{\small Schematic of the problem in phase space. The laminar-turbulent edge $\Sigma$ separates initial conditions that trigger turbulence (characterised by an initial energy $E_0$ greater than the critical value $E_c$) from those that decay back to the laminar state ($E_0<E_c$). The minimal seed (MS) is the point infinitesimally close to the boundary just capable of triggering turbulence.}
  \label{schematic-ms}
\end{figure}

The ultimate goal of these studies is to provide a characterisation of the basin of attraction of the laminar flow, i.e. the subset of initial conditions which asymptotically converge to the laminar state. However, these methods are impractical 
at finding the smallest possible solution capable of just kicking
the system away from the laminar state,
as they require a large number of simulations/experiments.
Recent developments have been achieved using variational methods to construct fully nonlinear optimisation problems that seek the minimal seed \citep{pringle-kerswell-2010, pringle-etal-2012,cherubini-etal-2012,duguet-etal-2013, cherubini-palma-2014}; see \citet{kerswell-2018} for a review. From a dynamical-systems point of view, the minimal seed represents the closest (in a chosen norm) point of approach of the laminar-turbulent boundary, or `edge', to the basic state in phase space, as shown in figure \ref{schematic-ms}. If transition is regarded as undesirable, such perturbation will be considered the `most dangerous' disturbance. Previous studies \citep[][referred to as PK10 and PWK12, respectively, throughout the paper]{pringle-kerswell-2010, pringle-etal-2012} have revealed important characteristics of the minimal seed, such as its fully-localised nature and its three-phase evolution consisting of the Orr mechanism, the oblique phase and the lift up, during which the flow gradually unwrap to give rise to a large, predominantly streamwise independent final state. However, a link between the critical initial energies of the minimal seed and those of disturbances that can typically be generated in a lab has not been provided yet. This will be the focus of the first part of the paper, with the outcomes being summarised in the key graph, figure \ref{fig-statistical-study}, where the scaling $E_c = E_c(Re)$ for the minimal seed and several forms of randomised initial conditions are compared.

\subsection{Control of pipe flows}
\label{intro:control}
Several different control strategies have been designed in the past fifty years to reduce the wall friction of fully turbulent flows \citep[refer, for example, to][for reviews]{lumley-blossey-1998,kasagi-etal-2009,quadrio-2011}. In light of the central role of streamwise vortices in the drag and shear stress production \citep[e.g.][]{kim-etal-1987, W97}, \citet{choi-etal-1994} proposed an `opposition control technique' aimed at actively counteracting vortices or selected velocity components to reduce the skin-friction drag on the wall. \citet{xu-etal-2002} applied suboptimal opposition control \citep{lee-etal-1998} to pipe flow at a wall Reynolds number $Re_{\tau} = 150$ and achieved drag reduction of approximately 13\% to 23\%. Both passive (e.g. riblets) or active (oscillations or generation of travelling waves) methods have been employed to inhibit the near wall turbulence creation. Drag reductions of 25\% to 40\% were obtained in fully turbulent pipe flows using spanwise wall oscillations \citep{choi-graham-1998,duggleby-etal-2007,quadrio-sibilla-2000, choi-etal-2002,zhou-ball-2008}. \citet{zhou-ball-2008} also considered streamwise oscillations but found this method to be ineffective. In their experimental and numerical study, \citet{auteri-etal-2010} were able to achieve 33\% drag reduction by imposing streamwise-modulated waves of spanwise velocity travelling forward in the streamiwse direction. \citet{WHC10} also found a possibility to reduce drag by forcing large scale streaks in pipe flow and reported a power saving up to 11\%.

More recently, large-scale control methods that completely relaminarise fully turbulent flows by manipulating the mean profile have been successfully employed \citep{hof-etal-2010, kuhnen-etal-2018a}. These methods target the mean shear in order to counteract its crucial role as energy source in near wall turbulence \citep{schoppa-hussain-2002}. Based on the observation that the streamwise vorticity of a turbulence puff is mainly produced at the trailing edge by the fast incoming flow, \citet{hof-etal-2010} developed both experimental and numerical methods to flatten the velocity profile at the upstream edge of the puff to intercept this mechanism and successfully relaminarise the puff. Their idea was further developed by \citet{kuhnen-etal-2018a}, who were able to achieve a complete and final collapse of turbulence by appropriate distortions of the mean profile, with the friction losses reduced by as much as 90\%. Compared to some of the other strategies presented so far (for example, the opposition control method, which requires a knowledge and detailed manipulation of the fully turbulent velocity field), this approach is much simpler to implement as it only requires a steady open-loop manipulation of the streamwise velocity component. Experimentally, full relaminarisation was obtained by either increasing the turbulence level by vigorously stirring the flow with rotors or via wall-normal injection of additional fluid through several small holes in the pipe wall or by accelerating the flow close to the wall via streamwise injection of fluid through an annular gap at the wall or by means of a movable pipe segment. The common and key ingredient to these relaminarisation techniques was a flattened streamwise velocity profile, i.e. a deceleration of the flow in the bulk region and/or an acceleration of the flow close to the wall. The important role of the mean-flow distortion was confirmed numerically by adding a global body force to the equations of motions such that the resulting velocity profile was more `plug-shaped'. The efficiency of the control mechanism was directly related to the suppression of the lift up mechanism \citep[reviewed recently by][]{brandt-2014}, measured by the linear transient growth. All disturbances schemes were shown to lead to a reduction of the linear transient growth, that is, the modified profile was shown to suppress the energy transfer from the mean flow to the streamwise vortices and to inhibit the streak-vortex interaction. 
 
Most of the literature pertaining to the control of shear flows is devoted to suppressing fully turbulent flow. However, delaying (or preventing) transition to turbulence, thus avoiding the worst of turbulence {\emph{in toto}}, is even more desirable. Nevertheless little literature is available on this subject. Suppressing the energy growth of initial perturbations to delay or prevent transition requires an understanding of how the basin of attraction of the laminar flow is modified in the presence of the control. So far, theoretical work has focused on investigating the sensitivity of the linearised Navier-Stokes equations around the laminar state in order to design suitable controls \citep{jovanovic-2008}. This approach has had some success in mitigating turbulence transition using both open-loop and feedback-based approaches \citep{kim-bewley-2007}. For example, \citet{hogberg-etal-2003} used direct numerical simulation to demonstrate that linear feedback control strategies can significantly expand the laminar state's basin of attraction of plane Couette flow for a range of Reynolds numbers. In channel flows, \citet{moarref-jovanovic-2010} performed a perturbation analysis in the wave amplitude of the linearised Navier-Stokes equations to `design' travelling waves which significantly reduce the sensitivity of the flow. However, as pointed out by \citet{bewley-2001}, due to the finite-amplitude nature of transition in shear flows, a fully-nonlinear approach is required to probe the sensitivity of the laminar state to finite-amplitude disturbances. In a proof-of-concept study, \cite{rabin-etal-2014} showed how an optimisation approach could be used to design a more nonlinearly stable plane Couette flow through manipulation of the boundary conditions. By spanwise oscillating one boundary (with amplitude $A$ and frequency $\omega$), these authors showed that $E_c$ could be increased by 40\% through judicious choice of $A$ and $\omega$.

Our study is motivated by a recent experimental observation \citep{kuhnen-etal-2018b} that manipulation of the flow in a pipe with a baffle can lead to full relaminarisation for flow rates up to 3 times ($Re=6000$) that for which turbulence typically appears in the presence of ambient perturbations ($Re>2000$). Our focus is on theoretically capturing the phenomenon observed in experiments so that the process can then be optimised.

%%%%%%%%%%%%%%%%%%%%%%%%%%%%%%%%%%%%%%%%%%%%%%%%%%%%%%%%%%%%%%%%%%%%%%%%%%%%%%%%%%%%%%%%%%%%%%%%%%%%%%%%%%%%%%%%%%%%%%%%%%%%%%%%%%%%%%%%%%%%%%%%%%%%%%%%%%%%%%%%%%%%%%%%%%
\section{Formulation}
\label{sec:formulation}
We consider the problem of constant mass-flux fluid flow through a straight cylindrical pipe of length $L$ and diameter $D$. The flow is described using cylindrical coordinates $\{r,\theta,z\}$ aligned with the pipe axis. Length scales are non-dimensionalised by the radius of the pipe $D/2$ and velocity components by the laminar centerline velocity $2\,\overline{W}$, where $\overline{W}$ is the constant bulk velocity. Unless otherwise specified, energies are given as `absolute energies', i.e. not scaled by the energy of the laminar flow in the same domain. We consider a perturbation $\mathbf{\tilde{u}}=\{\tilde{u},\tilde{v},\tilde{w}\}$ superimposed on the laminar Hagen-Poiseuille flow (HPF) $\mathcal{W}(r)\mathbf{\hat{z}}=(1-r^2)\mathbf{\hat{z}}$ so that the full velocity field is given by $\mathbf{u}_{tot}=\mathcal{W}(r)\mathbf{\hat{z}}+\mathbf{\tilde{u}}(r,\theta,z,t)$.
The problem is governed by the continuity and Navier-Stokes equations
\beq
\mathbf{NS}=\frac{\p \mathbf{\tilde{u}}}{\p t} + \mathcal{W}\frac{\p \mathbf{\tilde{u}}}{\p z} + \tilde{u}\mathcal{W}'\mathbf{\hat{z}} -\mathbf{\tilde{u}} \times \nabla \times \mathbf{\tilde{u}} + \nabla \tilde{p} - \frac{4\beta}{Re}\mathbf{\hat{z}}-\frac{1}{Re} \nabla^2 \mathbf{\tilde{u}} =0,\;\;\;\; \nabla \cdot \mathbf{\tilde{u}}=0\,,
\eeq
where the prime indicates total derivative, $Re=\overline{W}D/\nu$ is the Reynolds number and $\beta=\beta(\tilde{\mathbf{u}})$ is a correction to the pressure such that the mass flux remains constant. The parameter $1+\beta$ is an observed quantity in experiments and is defined as the ratio of the observed dissipation $\mathcal{D}$ \citep{popebook} (or pressure gradient $\langle \p p /\p z \rangle$) and the corresponding laminar value $\mathcal{D}_{lam}$ (or laminar pressure gradient $\langle \p p /\p z \rangle_{lam}$), namely
\begin{equation}
 1+\beta=\frac{\mathcal{D}}{\mathcal{D}_{lam}}=\frac{\langle \p p /\p z \rangle}{\langle \p p /\p z \rangle_{lam}} \, ,
\label{1plusbeta}
\end{equation}
where the angle brackets indicate the volume integral
\beq
\langle ... \rangle = \int_0^L \int _0^{2\pi}\int_0^1 ... r\mathrm{d}r\mathrm{d}\theta\mathrm{d}z\,.
\eeq
Periodic boundary conditions are imposed in the streamwise direction and no-slip/no-penetration conditions on the pipe wall.

The formulation of the nonlinear variational problem closely follows PWK12 and the reader is referred to their section 2 for a detailed explanation.
In its simplest form the problem can be stated as follows: among all (incompressible) initial conditions of a given perturbation energy $E_0$, we seek the disturbance that gives rise to the largest energy growth
\beq 
G(T,E_0)=\displaystyle \max_{E_0 } \frac{\langle \mathbf{u}(\mathbf{x},T)^2\rangle}{\langle\mathbf{u}(\mathbf{x},0)^2\rangle} \, .
\eeq
To find the minimal seed, the initial energy $E_0$ is gradually increased and the variational problem solved until the critical energy $E_c$ is reached where turbulence is just triggered. Ideally, for asymptotically long times $T=T_{opt}$, $G$ is expected to approach a step function in $E_0$, with the jump at the critical value $E_c$. 
In practice, two conjectures proposed by PWK12 are exploited. For asymptotically large $T_{opt}$, the initial energy $E_{fail}$ at which the algorithm first fails to converge corresponds to $E_c$ (conjecture 1) and the converged nonlinear optimal (NLOP) approaches the minimal seed as $E_0$ approaches $E_c$ (conjecture 2).

The calculations are carried out using the open source code \texttt{Openpipeflow} \citep{willis-2017}, with a variable $q$ discretised in the domain $\{r,\theta, z\}=[0,1]\times[0,2\pi]\times[0,2\pi/\alpha]$ using Fourier decomposition in the azimuthal and streamwise direction and finite difference in the radial direction, i.e.
\beq
q(r_n,\theta,z) = \sum_{k<|K|} \sum_{m<|M|} q_{n,k,m}e^{i(\alpha k z + m \theta)}\,,
\eeq
where $n=1,...,N$ and $\alpha$ is the streamwise wavenumber. The radial points are clustered close to the wall. For a pipe of length $5D$ at $Re=2400$, we use $N=64$,  $K=36$, $M=32$, and time step $\Delta t=0.01$, with the discretisation appropriately refined as the Reynolds number is increased to keep the resolution unaltered. Unless otherwise specified, throughout the paper we use $L = 5D$.
%%%%%%%%%%%%%%%%%%%%%%%%%%%%%%%%%%%%%%%%%%%%%%%%%%%%%%%%%%%%%%%%%%%%%%%%%%%%%%%%%%%%%%%%%%%%%%%%%%%%%%%%%%%%%%%%%%%%%%%%%%%%%%%%%%%%%%%%%%%%%%%%%%%%%%%%%%%%%%%%%%%%%%%%%%
\section{Results and discussion}
\label{sec:results}
%------------------------------------------------------------------------------------------------------------------------------------------------------------------------
\subsection{Robustness of the minimal seed}
\label{sec:robustness}
%------------------------------------------------------------------------------------------------------------------------------------------------------------------------
\subsubsection{Changes to the base flow}
As a preliminary study we look at minimal seed with a modified base flow. By adding a body force, the laminar base profile becomes more `plug-like', as in \citet{kuhnen-etal-2018a}. We aim to investigate how the basin of attraction of the laminar state is affected by changes of the base flow and if this modification affects the minimal seed, i.e. the fingertip of the edge shown schematically in figure \ref{schematic-ms}.

\begin{figure}
  \centering
%	\subfloat{\includegraphics[width=0.5 \textwidth]{./fig-forcing-profile.pdf}}
%	\subfloat{\includegraphics[width=0.5 \textwidth]{./fig-forcing.pdf}}
\includegraphics[width=1.0\textwidth]{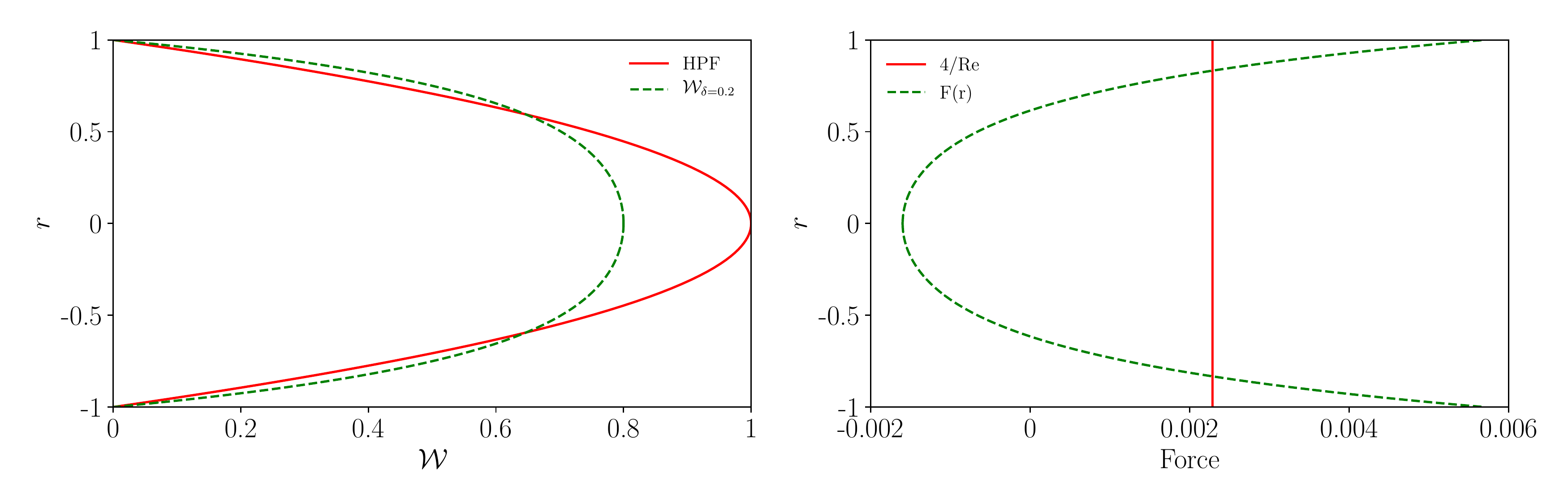}
  \caption{\small Left: comparison of the parabolic profile (red solid line) and the forced profile (green dashed line) of the laminar flow.  Right: the shape of the force chosen to decrease the centerline velocity of the laminar flow by 20\% at $Re=1750$. The force is negative (pointing upstream) in the centre and positive (pointing downstream) near the wall. For comparison, the pressure gradient in a laminar flow $|\partial p/ \partial z| =4/Re$ is also shown.}
  \label{fig-forced}
\end{figure}

Following \citet{kuhnen-etal-2018a} we use the following family of profiles for the base flow $\mathcal{W}(r;\delta)\mathbf{\hat{z}}$ (see their equation 19), where 
  \beq
     \mathcal{W}(r;\delta)=(1-\delta)\left[1 - \frac{\cosh(cr) - 1}{\cosh(c) - 1} \right]\,.
     \label{forc-base-prof}
  \eeq
The parameter $\delta$ is the centreline difference between the laminar profile and the target profile and $c$ is chosen by imposing the constant mass flux condition. The force $F = F(r)\hat{\mathbf{z}}$ required to generate such a target velocity profile is obtained by substituting $\mathcal{W}(r;\delta)$ in the Navier Stokes equations, i.e. $F(r):=-\hat{\mathbf{z}}\cdot\mathbf{NS}$. For example, the case with $\delta=0.2$ and $c=3.5935$ is shown in figure \ref{fig-forced} \citep[cfr][supplementary material]{hof-etal-2010,kuhnen-etal-2018a}. The forcing decelerates the flow in the central part and accelerates it near the wall while the mass flux is kept fixed.

The effect of the forcing is studied for the parameters corresponding to the works by PK10 and PWK12 as summarised in table \ref{table-1}. As discussed in PWK12, the choice of parameters of PK10 is not entirely appropriate: the Reynolds number is close to the first appearance of turbulent state, the target time is short and therefore the algorithm struggles to discern between conditions that relaminarise and conditions that trigger turbulence and in tightly constrained geometry the basin boundary is highly fractal. Nevertheless it is useful here to show the effect of the global forcing.

\renewcommand{\arraystretch}{1.3}
\setlength\tabcolsep{0.4cm}
\begin{table}
\centering
\begin{tabular}{c c c c c c }
 \vspace{0.2cm}
 Case &$Re$ & $\alpha$ &$L/D$ &$T_{opt}(D/\overline{W})$ &$N \times K \times M$\\

 PK10   &1750   &2     &0.5$\pi$ &21.35 &$60 \times 8 \times 16$ \\

 PWK12  &2400   &0.628 &5        &75   &$64 \times 36 \times 32$\\

\end{tabular}
\caption{Parameters and resolution for PK10 and PWK12 cases}
\label{table-1}
\end{table}

\begin{figure}
  \centering
%  \subfloat{\includegraphics[width=0.88 \textwidth]{./fig-cfr-GvsE0-PK10.pdf}}\\
%  \subfloat{\includegraphics[width=0.88 \textwidth]{./fig-cfr-GvsE0-PWK12.pdf}}
\includegraphics[width=0.8\textwidth]{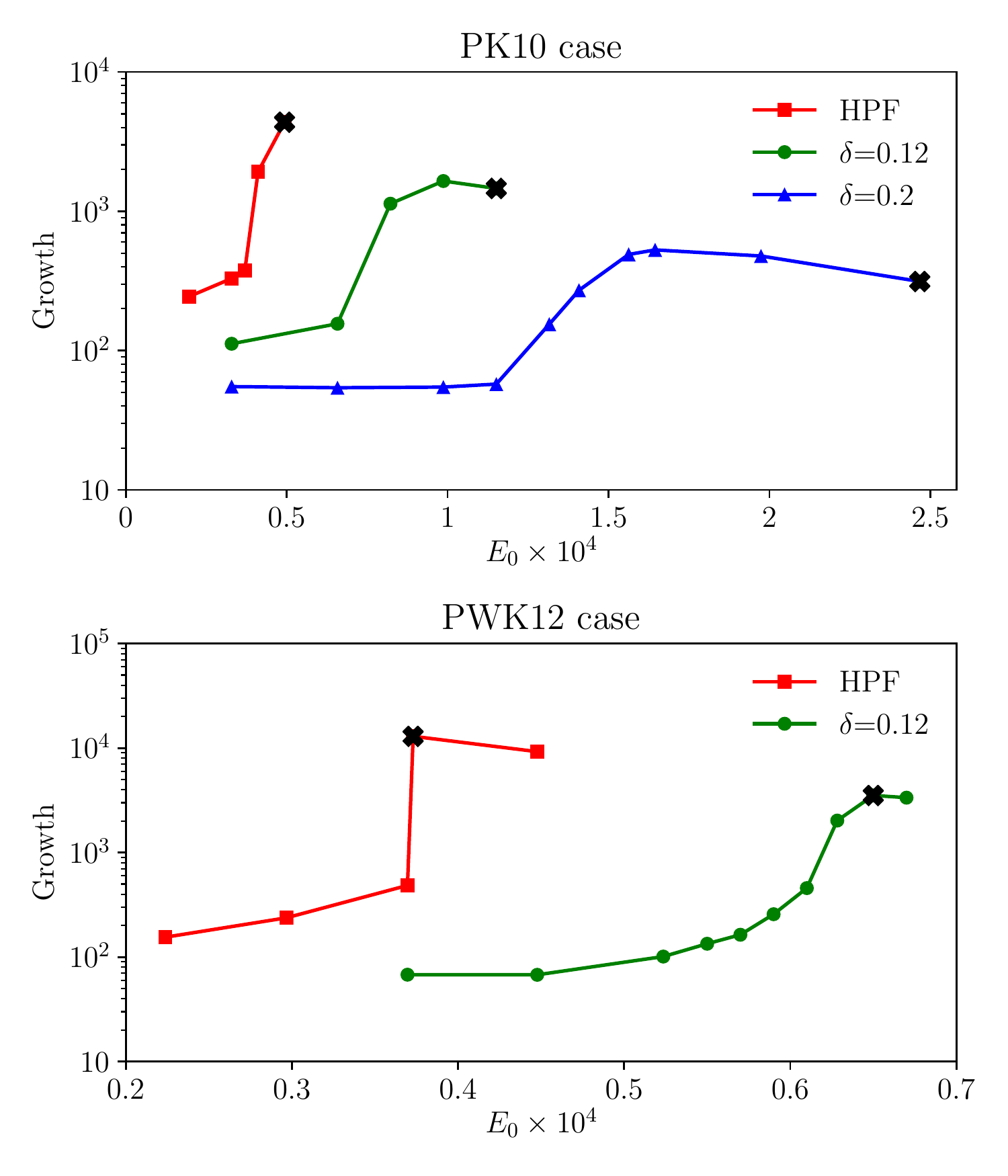}
  \caption{\small Comparison of $G$ vs $E_0$ with parabolic and forced base profile for PK10 and PWK12 cases (refer to table \ref{table-1}). The thick crosses indicate the initial energies $E_{fail}$ at which the optimisation algorithm first fails to converge.}
\label{fig-cfr-GvsE0}
\end{figure}

Figure \ref{fig-cfr-GvsE0} shows the maximum growth (at the target time $T_{opt}$) as a function of $E_0$ for the cases with forced and parabolic base profile. The initial energy is gradually increased until $E_c$ is reached.
Most of the data points have been verified by feeding the algorithm with at least two or three different initial conditions, for example a snapshot from a turbulent run, another NLOP at a lower $E_0$ or a turbulence-inducing initial conditions at a higher $E_0$. The initial energies $E_{fail}$ at which the optimisation algorithm first fails to converge are marked with a cross in figure \ref{fig-cfr-GvsE0}. According to the conjecture 1 of PWK12, these correspond to the critical initial energies $E_c$, where the edge touches the energy hypersurface at one velocity state. Due to the reasons mentioned above, in the PK10 case, especially for $\delta=0.2$, convergence is sometimes not clear and deteriorates (becomes slower and slower) as $E_0$ is increased and approaches $E_{fail}$. The last data points before $E_{fail}$, for both values of $\delta$, appear to show convergence, but there still remains some doubt even after running the algorithm for more than 1000 iterations. Furthermore, the perturbations corresponding to $E_0=E_{fail}$ decay immediately after $T_{opt}$ is reached, because at this low Reynolds number turbulence is intermittent and appears only in the form of decaying puffs.

For the PWK12 case, convergence is clearer than in the PK10 case due to the larger domain, longer integration time and higher Reynolds number. However, in the forced case at $E_0 = 6.28 \times 10^{-5}$ (last data point before $E_{fail}$) neither a smooth convergence nor a clear increase of the residual was obtained, even after 1000 iterations. Note also that in the unforced case, $G$ sharply increases when the critical initial energy is reached, while with a flattened base profile, the increase of $G$ is much more gradual, as this case behaves similarly to cases where the system is close to the marginal $Re$ (as in PK10 case discussed above).

\begin{figure}
  \centering 
\begin{tikzpicture}
\tikzstyle{longdashed}=                  [dash pattern=on 5pt off 3pt]
% Main figure
\node at (-13,0) {\hspace{-0.6cm}\includegraphics[width=0.6\textwidth]{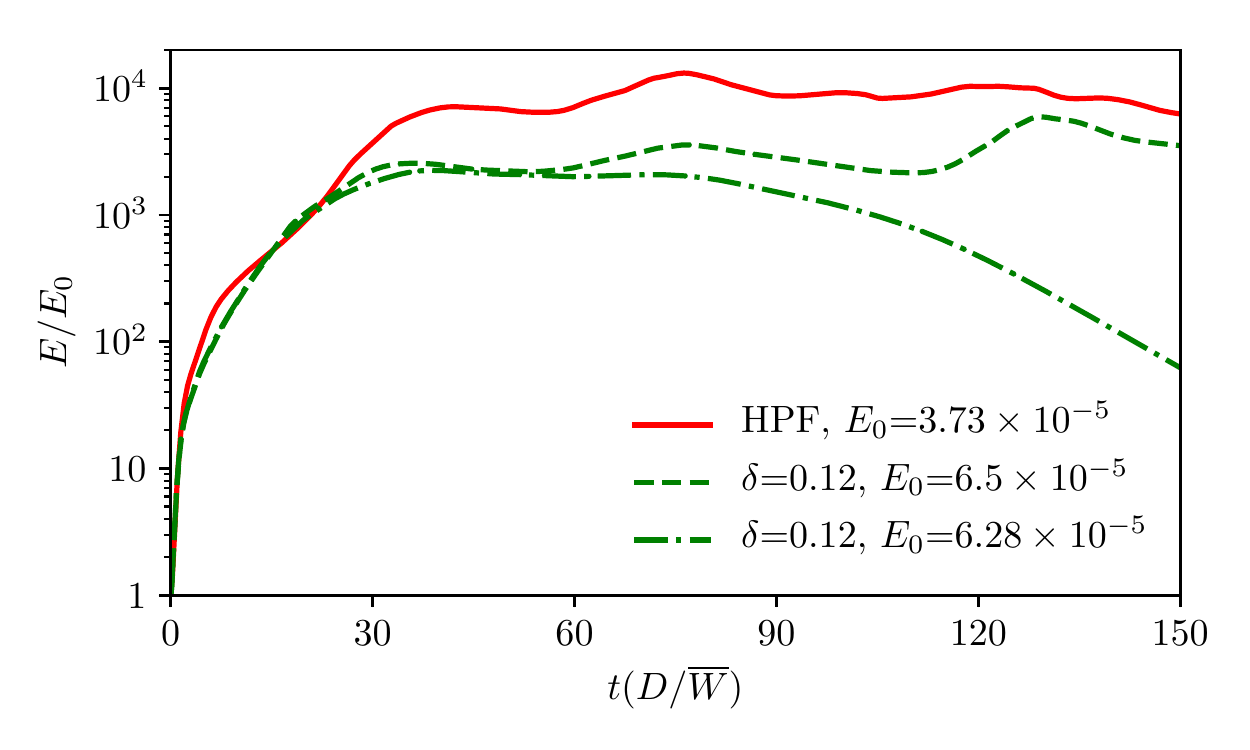}};
% Inset 1
\node[draw=red,thick, thick, line width=0.4mm]  at (-8,0.25) {\includegraphics[width=0.173\textwidth]{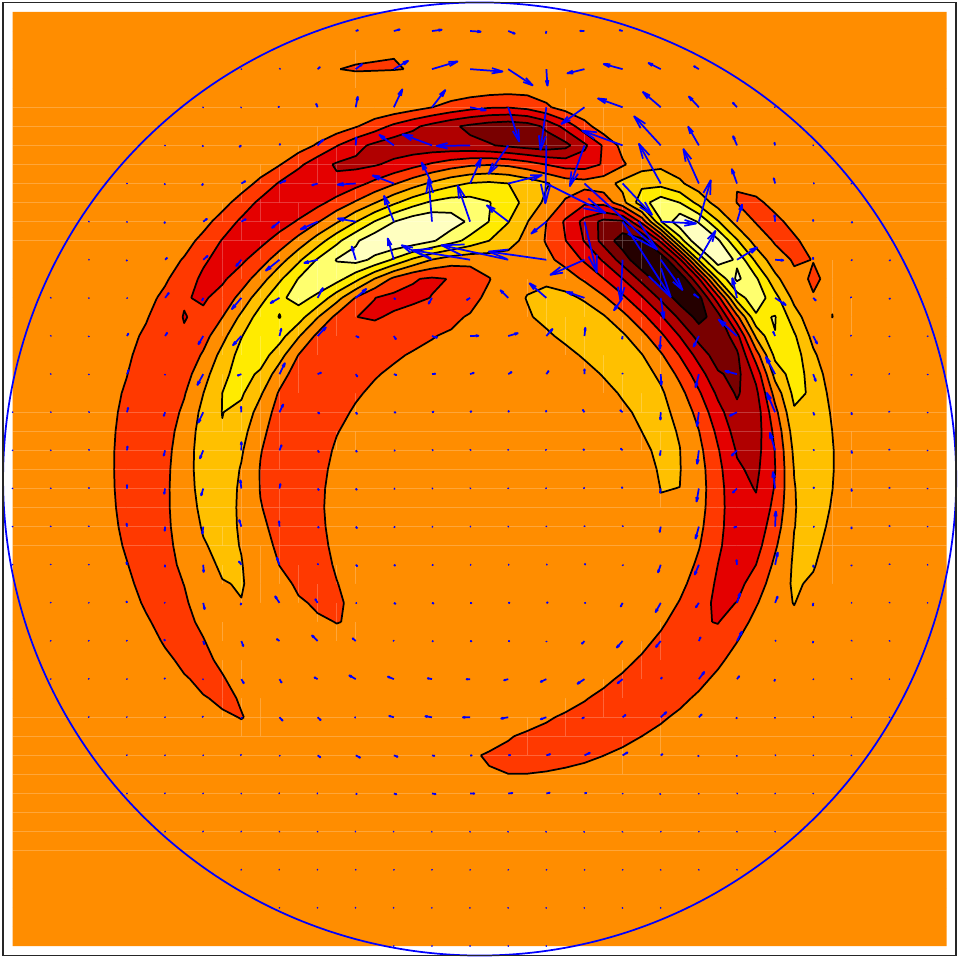}};
% Inset 2
\node[draw=green!50!black,thick, thick, line width=0.4mm, longdashed] at (-5.2,0.25) {\includegraphics[width=0.173\textwidth]{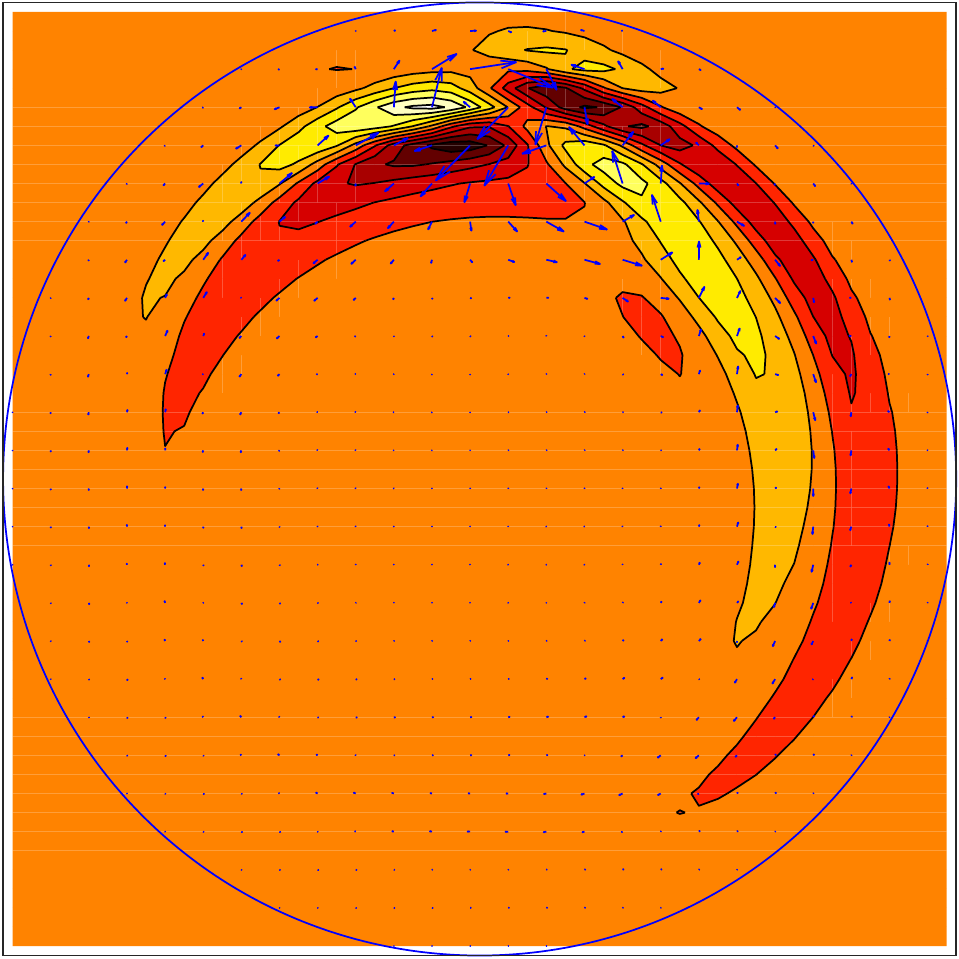}};

\end{tikzpicture}
  \caption{\small Left: time series of energy for the minimal seeds of PWK12 case (parameters provided in table \ref{table-1}) with forced (green dashed line) and unforced (red solid line) base profiles. In the forced case, the time evolution of the NLOP just before $E_{fail}$ is also shown (green dash-dotted line). Although in this case the convergence of the optimisation algorithm was not clear, the decay of the energy straight after $T_{opt}$ suggests that this disturbance does not lead to a turbulent episode. On the right: cross sections of the minimal seeds in the forced (green dashed border) and unforced (red solid border) cases. Contours indicate axial velocity perturbations (with the laminar flow subtracted off): white or light for positive perturbation, red or dark for negative, outside shade corresponds to zero. The arrows indicate cross-sectional velocities.}
  \label{fig-energy-beta-MS}
\end{figure}

Despite these convergence issues, figure \ref{fig-cfr-GvsE0} shows that by flattening the base profile, $E_c$ is moved towards higher values of the initial energy and the maximum growth reached at time $T_{opt}$ is decreased, i.e. the unforced curve $G=G(E_c)$ is shifted `down' and `right' as $\delta$ is increased. Therefore, the presence of the forcing expands the basin of attraction of the laminar base profile and reduces transient growth. For example, for PWK12 case, the critical energy of the minimal seed moves from $E_0=3.73 \times 10^{-5}$ to $E_0=6.5 \times 10^{-5}$ and for $E_0<E_{c}$ the NLOP of the forced case reaches less than half of the growth of the unforced case. The energy time series of the minimal seeds for the parabolic and forced cases are shown in figure \ref{fig-energy-beta-MS}. These initial conditions clearly lead to a turbulent episode which survives for at least double the optimisation time. The case $E_0=6.28 \times 10^{-5}$ for which convergence was critical is also shown. This initial state seems turbulent at the optimisation time but decays straight after.

%\begin{figure}
%  \centering 
%\begin{tikzpicture}
%\tikzstyle{longdashed}=                  [dash pattern=on 5pt off 3pt]
%% Main figure
%\node at (-8,0) {\hspace{-0.6cm}\includegraphics[width=0.88\textwidth]{./fig-PWK12-cfr-enrg-beta-MS.pdf}};
%% Inset 1
%\node[draw=red,thick, thick, line width=0.4mm]  at (-1.6,1.82) {\includegraphics[width=0.173\textwidth]{./fig-cs-minseed-0096.pdf}};
%% Inset 2
%\node[draw=green!50!black,thick, thick, line width=0.4mm, longdashed] at (-1.6,-0.89) {\includegraphics[width=0.173\textwidth]{./fig-cs-MS-forced-Re2400.pdf}};

%\end{tikzpicture}
%  \caption{\small Time series of energy for the minimal seeds of PWK12 case (parameters provided in table \ref{table-1}) with forced (green dashed line) and unforced (red solid line) base profiles. In the forced case, the time evolution of the NLOP just before $E_{fail}$ is also shown (green dash-dotted line). Although in this case the convergence of the optimisation algorithm was not clear, the decay of the energy straight after $T_{opt}$ suggests that this disturbance does not lead to a turbulent episode. On the right: cross sections of the minimal seeds in the forced (green dashed border) and unforced (red solid border) cases.}
%  \label{fig-energy-beta-MS}
%\end{figure}

Despite the critical initial energy being significantly increased with a flattened base profile, the fully localised structure of the minimal seed remains largely unchanged, as shown in the cross sections of figure \ref{fig-energy-beta-MS}. Therefore, the structure of the minimal seed is found to be fairly robust to changes to the base flow. This is different from \cite{rabin-etal-2014}'s study of oscillated plane-Couette flow where, instead, qualitative changes in the structure of the minimal seed are found as compared to the unoscillated case. In their study, however, the basic fluid response in the presence of spanwise oscillations becomes time dependent through the additional spanwise component (refer to their equation 2.1), while in our case only the shape of the laminar flow profile is modified, its dimension (1-D) and dependencies (only radial) remain unchanged.
%------------------------------------------------------------------------------------------------------------------------------------------------------------------------
\subsubsection{Filtering}
\begin{figure}
  \centering 
\begin{tikzpicture}

% Main figure
\node at (0,0) {\includegraphics[width=1.0\textwidth]{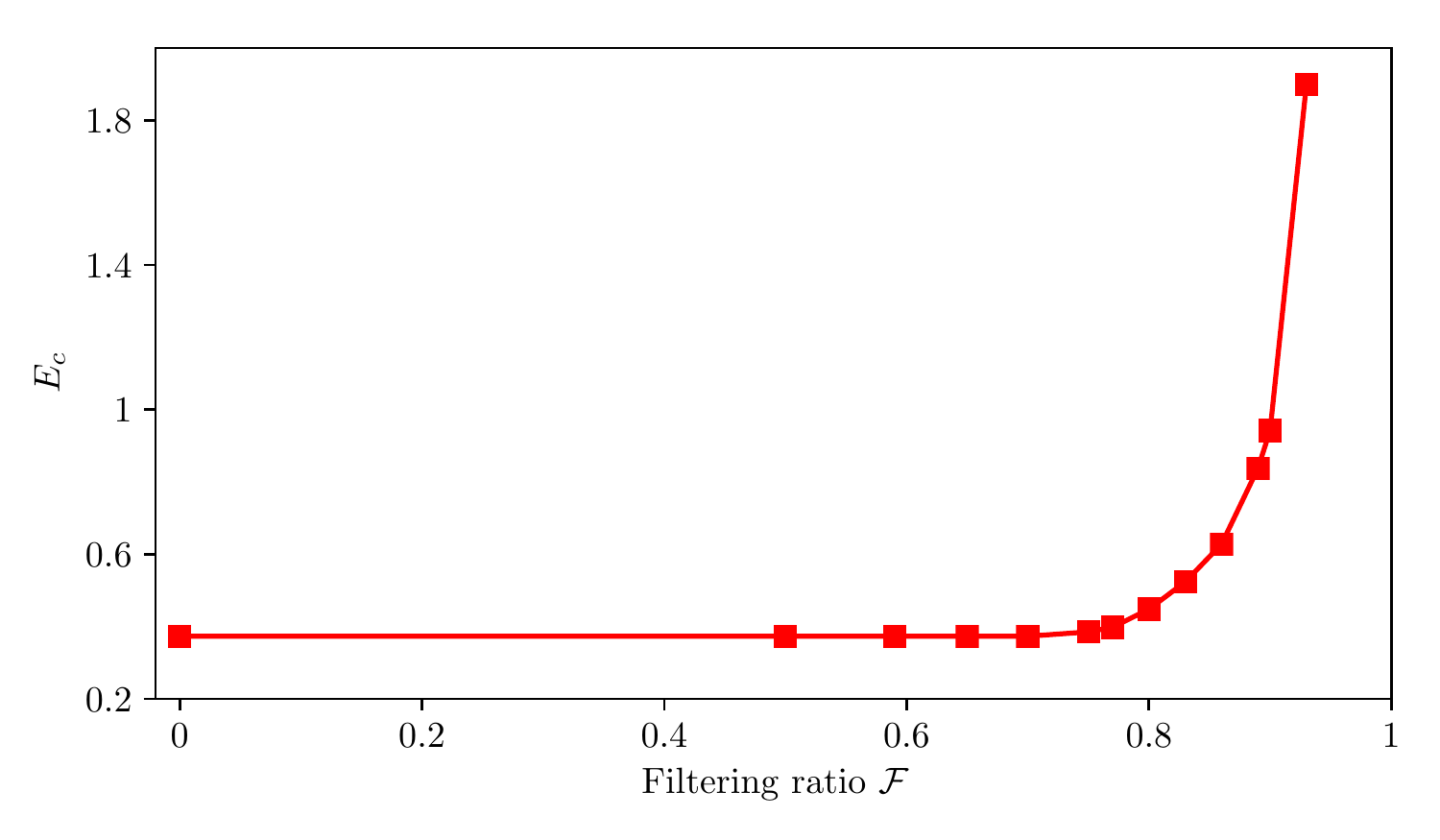}};
\node[black] at (-4.9,3.65) {$\times 10^{-4}$};
% Inset 1
\node (c1) at (-3.5,1) {\includegraphics[width=0.2\textwidth]{./fig-cs-minseed-0096.pdf}};
\draw[black,thick,->] (c1.south) -- (-4.98,-1.88);
\draw[black] (-5.12,-2.05) circle (0.2cm);

% Inset 2
\node (c2) at (-0.65,1) {\includegraphics[width=0.2\textwidth]{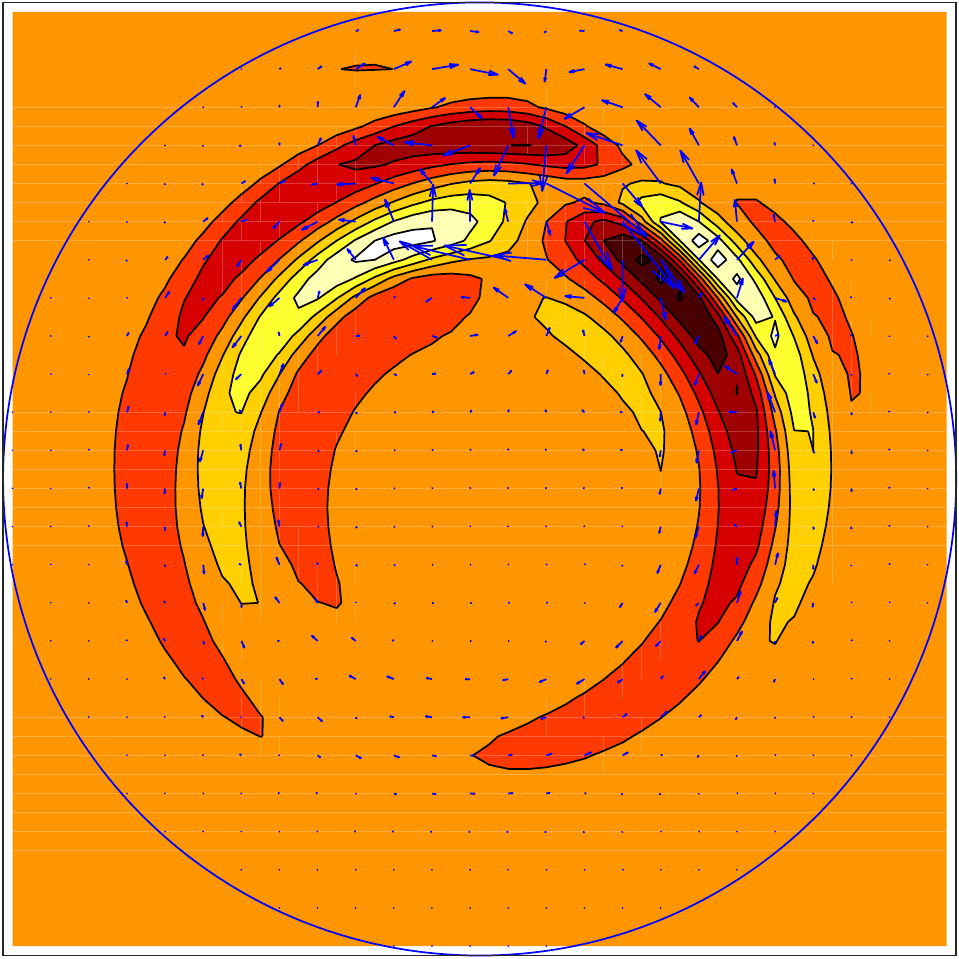}};
\draw[black,thick,->] (c2.south) -- (2.72,-1.9);
\draw[black] (2.9,-2.05) circle (0.2cm);

% Inset 3
\node (c3) at (2.2,1) {\includegraphics[width=0.2\textwidth]{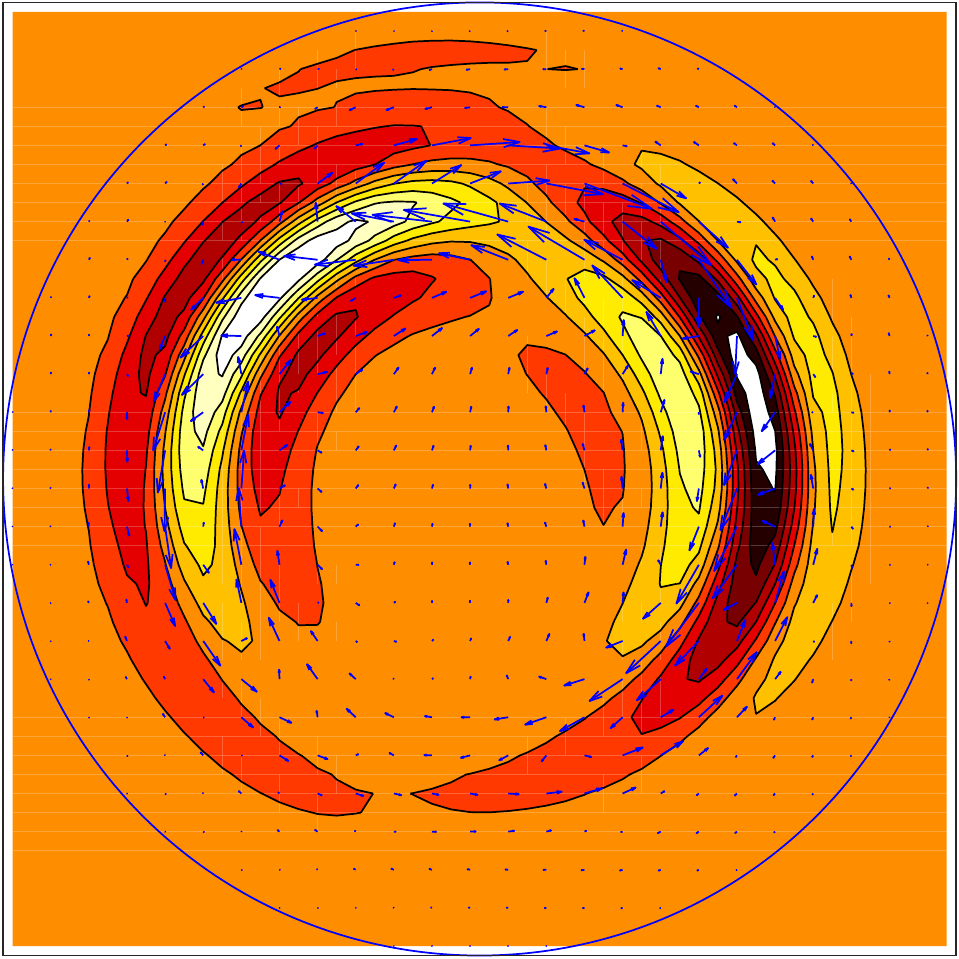}};
\draw[black,thick,->] (c3.east) -- (4.98,0);
\draw[black] (5.18,-0.1) circle (0.2cm);
\end{tikzpicture}
  \caption{\small Critical initial energy as a function of the filtering ratio $\mathcal{F}$ for the unforced PWK12 case (parameters provided in table \ref{table-1}). Insets: cross sections of the minimal seeds at different values of $\mathcal{F} = 0$, $0.75$ and $0.93$. Contours indicate streamwise velocity beyond the laminar flow (white or light for positive, red or dark for negative) and arrows indicate cross-sectional velocities.}
\label{fig-filtering}
\end{figure}

In order to assess how robust the minimal is to smoothening,
% we apply a spectral filter to the minimal seed found in the (unforced) PWK12 case, i.e. 
for the set of parameters corresponding to the (unforced) PWK12 case we perform the full optimisation procedure over perturbations in a lower dimension space, namely at each iteration we
 project the initial condition onto a subspace where only the first $K_f\times M_f$ wavenumbers are retained. Note that the resolution in the forward and backward steps is unchanged (i.e. $K=36$, $M=32$). As a measure of how much we are truncating we introduce the filtering ratio $\mathcal{F}$ defined as
\beq
\mathcal{F}=1-\sqrt{\frac{K_f}{K}\frac{M_f}{M}}\,.
\eeq
Figure \ref{fig-filtering} shows the critical energy of the minimal seed as a function of the filtering ratio, where $\mathcal{F}=0$ means no filtering (i.e. fully resolved minimal seed) and $\mathcal{F}=1$ would imply that no perturbation remains. Note that, for a fixed value of $\mathcal{F}$, $K_f$ and $M_f$ are chosen so that the corresponding ratios $K_f/K$ and $M_f/M$ are as close as possible, i.e. to avoid cases where the filtering in one direction is much higher than in the other direction. The critical initial energy of the minimal seed remains almost unchanged for $\mathcal{F}\le0.75$, that is, when only retaining 25\% of the modes. 
This is quite surprising because, since we are restricting the initial condition to a subset of the energy hypersurface by adding the filtering constraint, one would expect the minimal seed to occur at a higher initial energy. Our results thus suggest that the edge is locally quite flat at the minimal seed. As $\mathcal{F}$ is further increased, larger initial energies are needed in order to trigger turbulence. For example, for $\mathcal{F}=0.93$ (i.e. only the first $3\times2$ modes retained) $E_c$ is almost an order of magnitude larger than in the fully resolved case. For values of $\mathcal{F}$ larger than this (for example we have tested $\mathcal{F}=0.96$ where only $2\times$1 modes are retained) it seems that it is not possible to trigger transition. As shown by the cross-sections in the insets of figure \ref{fig-filtering}, the structure of the minimal seed remains almost unchanged when the filter is applied, even for cases where the minimal seed occurs at larger $E_c$ than in the unfiltered case (refer to the last inset to the right).

This study shows that the minimal seed is robust to quite severe spectral filtering, i.e. the small-scale structure of the minimal seed is not important. 
It is possible that radial filtering may have a more significant effect on the energy thresholds, but the structure remain essentially similar, thus suggesting that truncation is not a simple route towards a new set of more `typical' initial conditions. Conversely, it shows that the structure does not have to be perfectly formed to be the optimal. From this it seems reasonable that the minimal seed might be realised in a lab, but whether it would be `naturally' realised among `ambient' disturbances remains unclear at this stage.

%EM COMMENTED THE FOLLOWING
%So far we have shown that the minimal seed is fairly robust to changes to the base flow and to spectral filtering. However, robustness is only a `necessary but not sufficient' condition for the minimal seed to be a typical disturbance generated in a lab. In other words, if it were not robust, it would mean that the chances for any real perturbation to look like the minimal seed would be practically nil, but at the same time, this does not imply that the minimal is representative of typical disturbances generated in a lab. 
%------------------------------------------------------------------------------------------------------------------------------------------------------------------------
\subsection{Statistical study of transition to turbulence}
\label{sec:statistical}
To establish whether the minimal seed may be used to model typical ambient perturbations, we compare the critical initial energy of the minimal seed with that of a random disturbance. The latter is generated by scattering energy randomly over a subset of wavemumbers $K \times M = 12\times 9$. At $Re=2400$ this truncation corresponds to $\mathcal{F}\approx 0.7$,
before the rapid increase in $E_c$ (figure \ref{fig-filtering}), where the minimal seed could still be captured quite faithfully. An arbitrary complex amplitude $A_n$ is generated for each of the spectral modes in the chosen subset and the radial dependence introduced in the form $\sum_{n=1}^{N_s} A_n r \sin(n\pi r)$ with $N_s=5$. The projection scheme integrated into the time stepping algorithm is employed to ensure that the initial condition is solenoidal. These initial conditions are fed into DNS with time integration $T=125(D/\overline{W})$ (almost double the target time used in the adjoint algorithm). Five Reynolds number are considered: $Re=2400$, 3500, 5000, 7000 and 10000, with the numerical resolution appropriately increased for increasing Reynolds number. The same subset of wavenumbers used at $Re=2400$ is employed for all the other Reynolds number considered. For each $Re$ we consider 10 to 12 distinct random initial conditions and for each of them we gradually increase $E_0$ until turbulence is triggered and sustained for a time $T(D/\overline{W})$. The criterion to check for relaminarisation is $E_{3d}<$$10^{-8}$, where $E_{3d}$ is the energy associated with the streamwise dependent modes only, as this quantity decays very rapidly when the flow relaminarises. 

An analogous statistical study is carried out using 10 different snapshots from a turbulent run as initial condition (in a similar fashion to \citet{schneider-eckhardt-2008}). Both the random and snapshot initial conditions are {\emph {global}} disturbances, while the minimal seed is fully localised. A third set of random {\emph {localised}} initial conditions is obtained by scattering energy randomly over the wavenumbers as above, then by multiplying this global disturbance by a smooth spatial windowing function $\mathcal{B}(z)$ and $\mathcal{B}(\theta)$ so that the disturbance occupies only a 1/5 and a 1/3 of the streamwise and azimuthal domain, respectively. For the range of Reynolds numbers considered here we do not observe a strong localisation of the minimal seed in the radial direction and therefore we do not localise the random disturbance in this direction. The smoothing function is defined using equation 8 of \citet{yudhistira-skote-2011}. For example, for the localisation in the streamwise direction (analogous in the azimuthal direction for $\mathcal{B}(\theta)$):
\beq
\mathcal{B}(z)=g\left(\frac{z-z_{start}}{\Delta z_{rise}}\right) - g\left(\frac{z-z_{end}}{\Delta z_{fall}}+1\right)\,,
\label{smoothing-function}
\eeq
with
\[
 g(z^*) = 
  \begin{cases} 
   0 & \text{if } z^* \geq 0 \\
   \left\{1+\exp[1/(z^*-1)+1/z^*]\right\}^{-1} & \text{if } 0<z^*<1 \\

   1       & \text{if } z^* < 0
  \end{cases}
\]
where $z_{start}$ and $z_{end}$ indicate the spatial extent over which the disturbance is non-zero, and $\Delta z_{rise}$ and $\Delta z_{fall}$ are the rise and fall distances of the disturbance. 
%EM COMMENTED THE FOLLOWING
%These data are shown as blue crosses in figure \ref{fig-statistical-study} and we can notice that with localisation we are able to lower the critical energy by almost an order of magnitude.
\begin{figure}

  \centering 

\begin{tikzpicture}

\node at (0,0) {\includegraphics[width=1.0\textwidth]{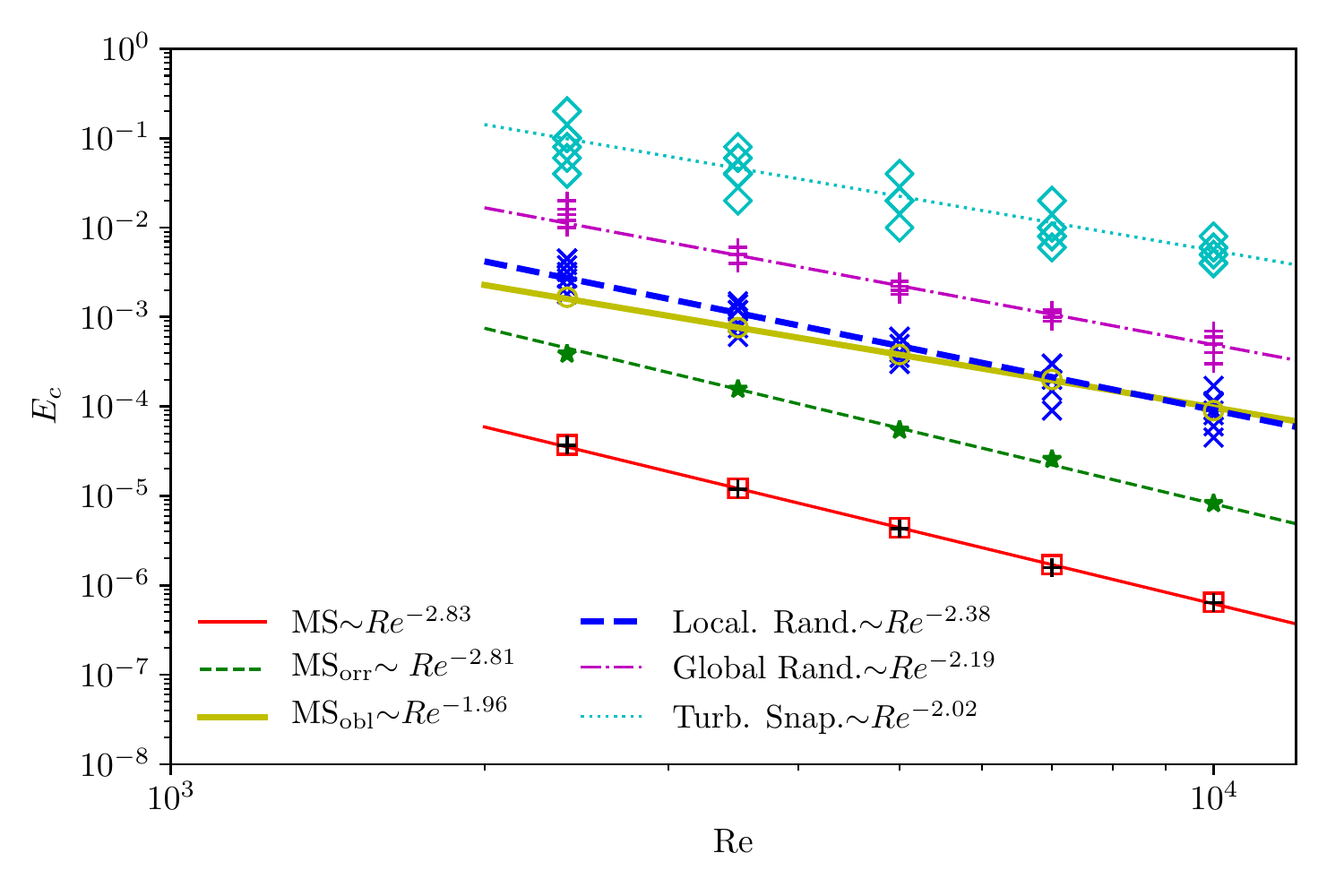}};

\tikzstyle{longdashed}=                  [dash pattern=on 6pt off 3pt]
\tikzstyle{longdashdotted}=              [dash pattern=on 6pt off 2pt on \the\pgflinewidth off 2pt]
\definecolor{magenta}{rgb}{1,0,1}  
\node[draw=magenta!75!black,thick,thick,longdashdotted] at (-3.65,2.7) {\includegraphics[width=0.15\textwidth]{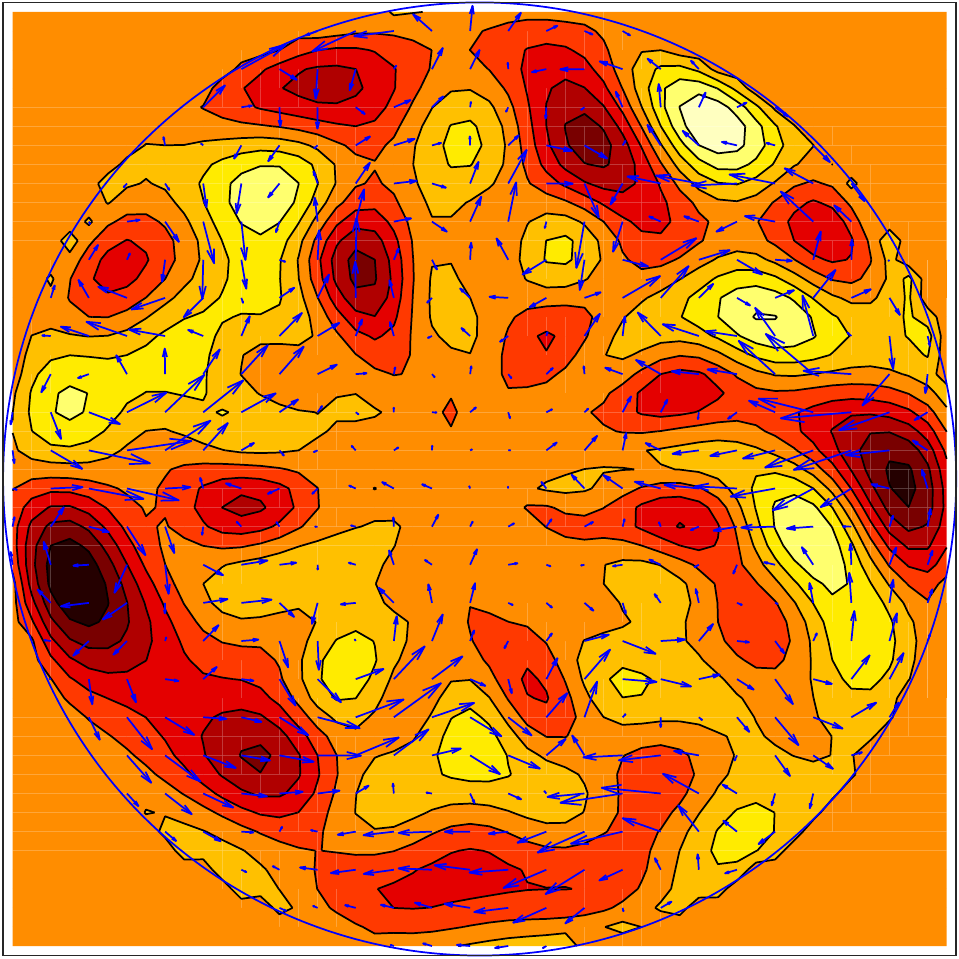}};
\node[draw=blue,line width=0.5mm,longdashed] at (-3.65,0.3) {\includegraphics[width=0.15\textwidth]{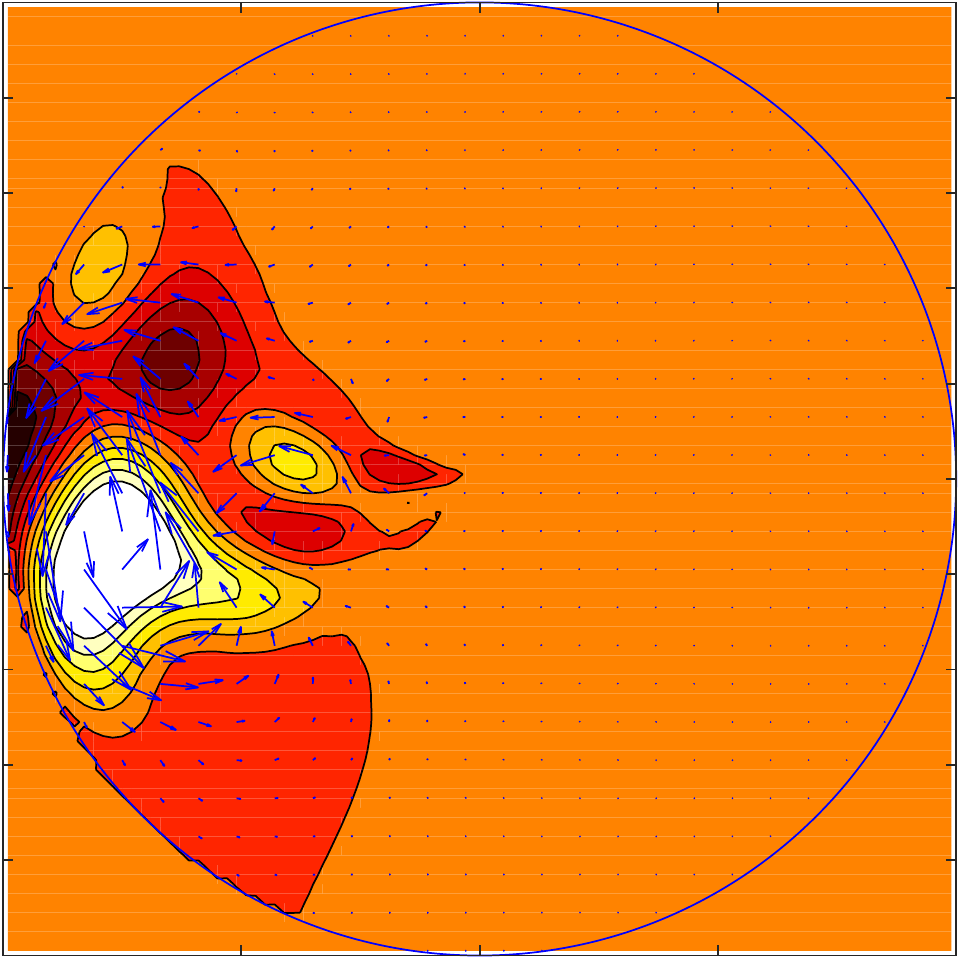}};

\end{tikzpicture}

  \caption{\small Energy thresholds $E_c$ vs $Re=2400$, 3500, 5000, 7000 and 10000 for different types of initial conditions: turbulent snapshots (cyan diamonds), artificially generated global (purple plusses) or localised (blue crosses) random fields and the minimal seed (black plus symbols correspond to the largest tested value of $E_0$ below which transition never occurs, while the red squares correspond to the smallest tested value of $E_0$ for which transition occurs at least once). The green stars and the yellow circles are the energies of the minimal seed after the Orr and oblique phases, respectively. The lines represent the power law scalings obtained by fitting the last four data points. Insets: cross sections of a typical global (purple dash-dotted border) and localised (dashed blue border) random initial condition. Contours indicate streamwise velocity beyond the laminar flow (white or light for positive, red or dark for negative) and arrows indicate cross-sectional velocities.}

\label{fig-statistical-study}

\end{figure}
 
The critical initial energies of the minimal seeds are found with a two-digit accuracy for all the Reynolds numbers considered, except for $Re=10000$ for which only one-digit accuracy is reached due to large computational cost of the simulations. The curves $E_c=Re^{-\gamma}$ obtained by fitting 
critical energies found for
the turbulent snapshot, global and localised random 
sets of data points and for the minimal seed are shown in figure \ref{fig-statistical-study}. The data points at $Re=2400$ are not used to obtain the fittings as turbulence can still be transient at this relatively low Reynolds number. A power-law exponent $\gamma=2.83$ for the minimal threshold energy is obtained.
%EM COMMENTED THE FOLLOWING
%A power law scaling $E_c=Re^{-\gamma}$ of the minimal threshold energy emerges, with $\gamma=2.83$.
 This result matches very well the exponent $\gamma\approx 2.8$ obtained experimentally by \citet{peixinho-mullin-2007} and later confirmed numerically by \citet{mellibovsky-meseguer-2009} using `push-pull' perturbations with constant flow rate. Minimal energies for transition to turbulence were also calculated by \citet{duguet-etal-2013} for plane Couette flow (refer to their figure 3), by \citet{cherubini-etal-2015} for the asymptotic suction boundary layer (refer to their figure 9) and, very recently, by Huang {\emph {et al.}} (personal communication) for plane channel flow. Our exponent is close to the $\gamma \approx 2.7$ obtained by \citet{duguet-etal-2013} for plane Couette flow and $\gamma \approx 3$ found by Huang {\emph {et al.}} for plane Poiseuille flow. Furthermore, our study is reminiscent of figure 19 in \citet{RSBH98}, where the streamwise-vortices and oblique-wave transition scenarios for plane channel flows are compared to two-dimensional linear optimals and noise, although the power-law exponents are not reported in their study.
The minimal seed curve $E_c=Re^{-2.83}$ is steeper than those pertaining to all the other forms of disturbances considered. The power law exponent $\gamma\approx 2$ for the turbulent snapshot recalls the scaling found by \citet{HJM03} and \citet{peixinho-mullin-2007} in their experiments using small impulsive jets to perturb the flow. For the random global and localised disturbances, larger exponents than for the turbulent snapshot are obtained, i.e. $\gamma=2.19$ and $2.38$, respectively (refer to table \ref{table-gamma} for a summary of the different exponents $\gamma$ discussed above). Most noticeably, the critical energy of the minimal seed is almost three orders of magnitude lower than that of the global disturbances considered here. With localisation of the random disturbances, the critical energy drops by almost an order of magnitude with respect to the global initial conditions, but it is still significantly larger than that of the minimal seed. 
\setlength\tabcolsep{0.3cm}
\begin{table}
\centering
\begin{tabular}{c c c c  }
 \vspace{0.2cm}
 Source & Geometry &Disturbance &Energy exponent $\gamma$ \\
\hline
 Present study   &HPF  &Turb. snapshot     &2.02 \\
 Present study   &HPF  &Global random          &2.19 \\
 Present study   &HPF  &Localised random       &2.38 \\
 Present study   &HPF  &Minimal seed       &2.83 \\
 \citet{HJM03} &HPF &Impulsive jets &$ 2$\\
 \makecell{\citet{peixinho-mullin-2007} \\ \citet{mellibovsky-meseguer-2009}} &HPF &\makecell{`Push-pull' with \\ const. mass flux}  &$ 2.8$\\
\citet{duguet-etal-2013} &PCF &Minimal seed  &$2.7$\\
\citet{cherubini-etal-2015} &ASBL &Minimal Seed &2\\
Huang {\emph {et al.}} &PPF &Minimal Seed &3

\end{tabular}
\caption{Power-law scalings $E_c=Re^{-\gamma}$ obtained in the present study and in the literature for different types of disturbances and in different geometries (HPF: Hagen-Poiseuille flow, PCF: plane Couette flow, ASBL: asymptotic suction boundary layer, PPF: plane Poiseuille flow).} 
\label{table-gamma}
\end{table}

Although the minimal seed has a very peculiar structure, which might be unlikely to be generated in a laboratory, it evolves to a structure that looks much more familiar to `natural' disturbance over a relatively short time scale corresponding to the Orr and oblique-wave phases.
The end of the Orr phase is identified by analysing the time evolution of the NLOP close to $t=0$: the streaks that are initially tightly layered and inclined back into the oncoming flow, are tilted away from the wall by the Orr mechanism and slightly separated (refer, for example, to figure 1 of PWK12). As observed in PK10 and PWK12 in the case of low Reynolds number, we find that the Orr phase occurs in a very short time scale ($0 < t(D/\overline{W}) \lesssim 0.5-1$) and gives rise to an initial spurt of energy growth. By the end of the Orr phase, the helical modes starts to become dominant, thus signalling that the oblique-wave mechanism has come into play.
 In PK10 the end of the oblique-wave phase was clearly signalled by the `shoulder' in the time evolution of energy (refer to their figure 1) and the corresponding rapid decay of $E_{3d}$ after reaching a peak. For the present choice of $L=5D$, which is approximately three times longer than 
that in PK10, the `shoulder' is not clearly distinguishable and $E_{3d}$ does not decay straight after the peak because of the longer length-scale modes. However, by defining a three-dimensional energy $\tilde{E}_{3d}=E_{3d}(k>3)$ that includes modes $k>3$, we are able to observe again the `shoulder' and the rapid decay of $\tilde{E}_{3d}$, in correspondence of which we identify the end of the oblique phase ($0.5-1 \lesssim t(D/\overline{W})  \lesssim 2.5-3.5$). Both the time scales of the Orr and oblique phases are found to be similar to the PK10 case and to remain almost unchanged with Reynolds number. 
%the streaks that are initially tightly layered and inclined back into the oncoming flow, are tilted away from the wall and slightly separated, as shown in figure 1 of PWK12. We can thus identify the end of the Orr phase by analysing the time evolution of the NLOP
% A similar time evolution of the NLOP is found in the present cases and therefore it is plausible to assume that also the Orr phase time scale remains approximately unchanged with different $Re$.
If we compare the energies of the minimal seed after the Orr and oblique phases to the random localised initial conditions (refer to figure \ref{fig-statistical-study}) then the gap is reduced to less than an order of magnitude in the former case and, most noticeably, to practically zero in the latter case.

Comparing the scalings of the critical initial energy of the minimal seed and of its energy at the end of the Orr phase suggests that the growth produced via the Orr mechanism is independent of the Reynolds number, as expected due to the inviscid nature of the Orr process. The growth produced via the oblique-wave mechanism is almost of $\mathcal{O}(Re)$.  
While the growth factor of $\mathcal{O}(Re^{2})$ for the maximum linear transient growth due to the lift-up mechanism is well documented \citep[e.g][]{schmid-henningson-2012}, to the best of our knowledge, this is the first time that scaling laws are obtained numerically for the Orr and oblique-wave mechanisms.

\begin{figure}
  \centering 
\includegraphics[width=1.0\textwidth]{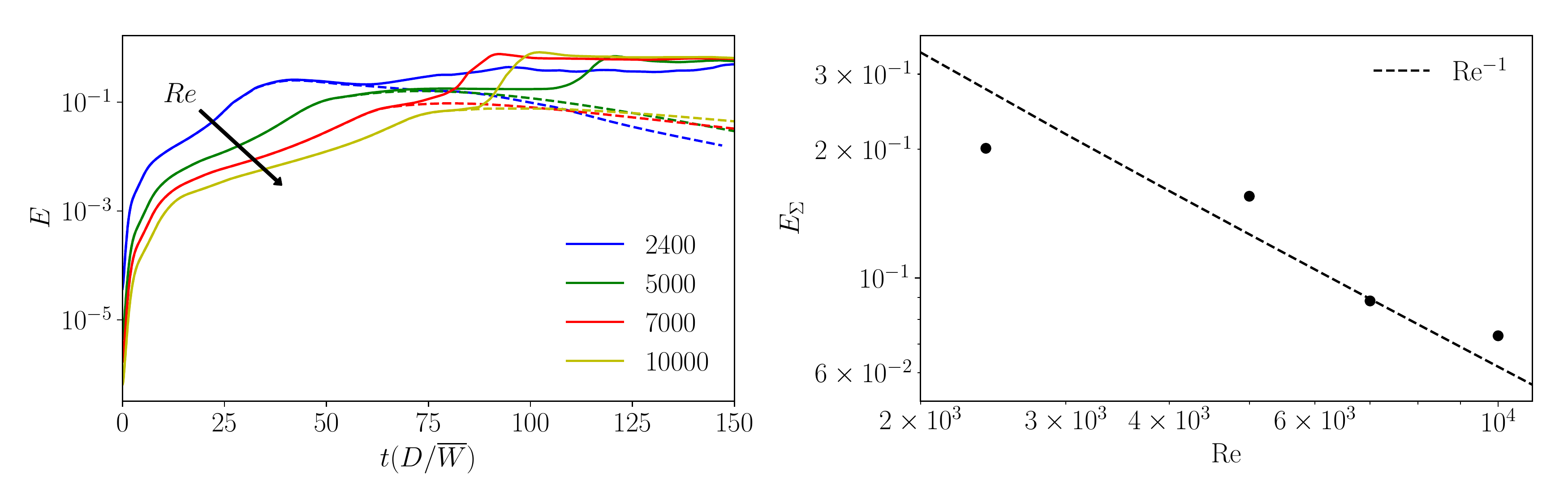}
 \caption{\small Left: trajectories close to the edge (solid lines: `just above', dashed line: `just below') at different Reynolds numbers in the range $2400 - 10000$. Right: energy of the edge as a function of $Re$, calculated as the point where the normalised difference between the laminar and turbulent trajectories becomes greater than 5\%. The dashed curve $Re^{-1}$ is added to show that our scaling assumption is reasonable, it is not calculated by fitting the data.}
  \label{fig-ecs-scaling}
\end{figure}

As shown by PWK12, the NLOP tracks the laminar-turbulent boundary $\Sigma$ before either relaminarising or triggering turbulence. The two bracketing cases shown in figure \ref {fig-statistical-study} as black plusses (relaminarising disturbances) and red squares (turbulence-inducing disturbances) are further refined with bisections until the difference in the initial energies is less than 0.005\%. These refined bracketing trajectories are shown in figure \ref{fig-ecs-scaling} for the range of Reynolds numbers considered and provide evidence of an edge tracked by the minimal seed. Our data (refer to figure \ref{fig-ecs-scaling}) suggest that the energy of an edge state $E_{\Sigma}$
decreases with increasing $Re$, approximately as $Re^{-1}$. 

\begin{figure}
    \centering
	\includegraphics[width=0.3\textwidth]{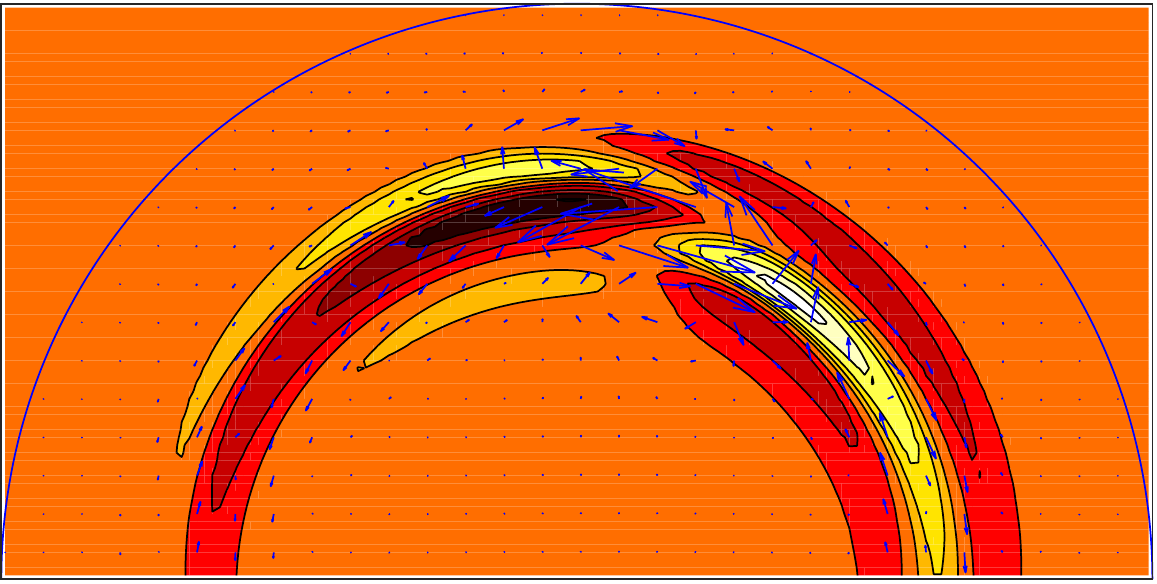}
	\includegraphics[width=0.3\textwidth]{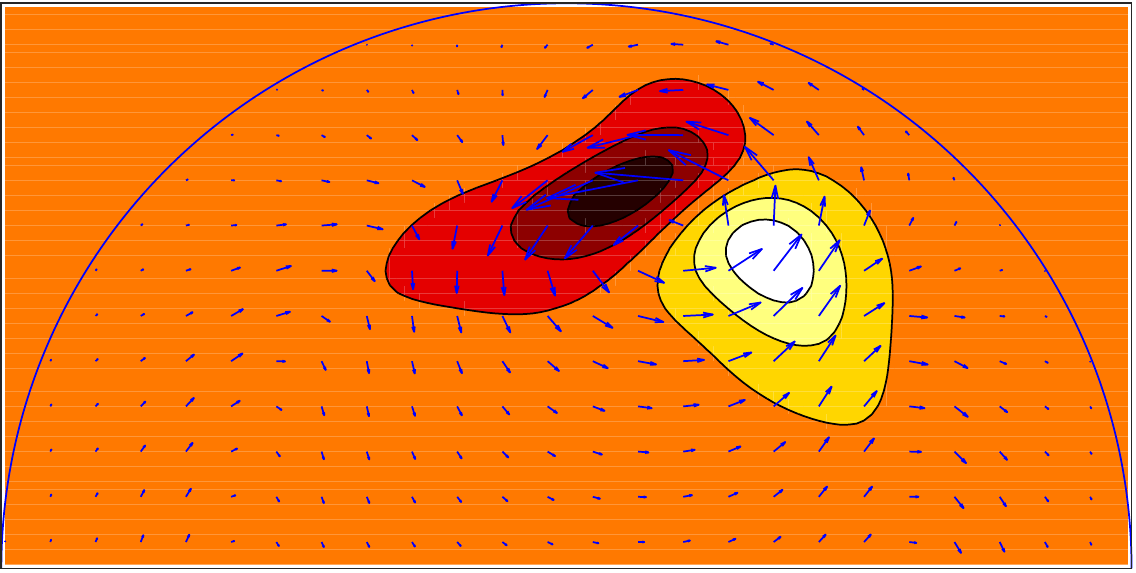}
	\includegraphics[width=0.3\textwidth]{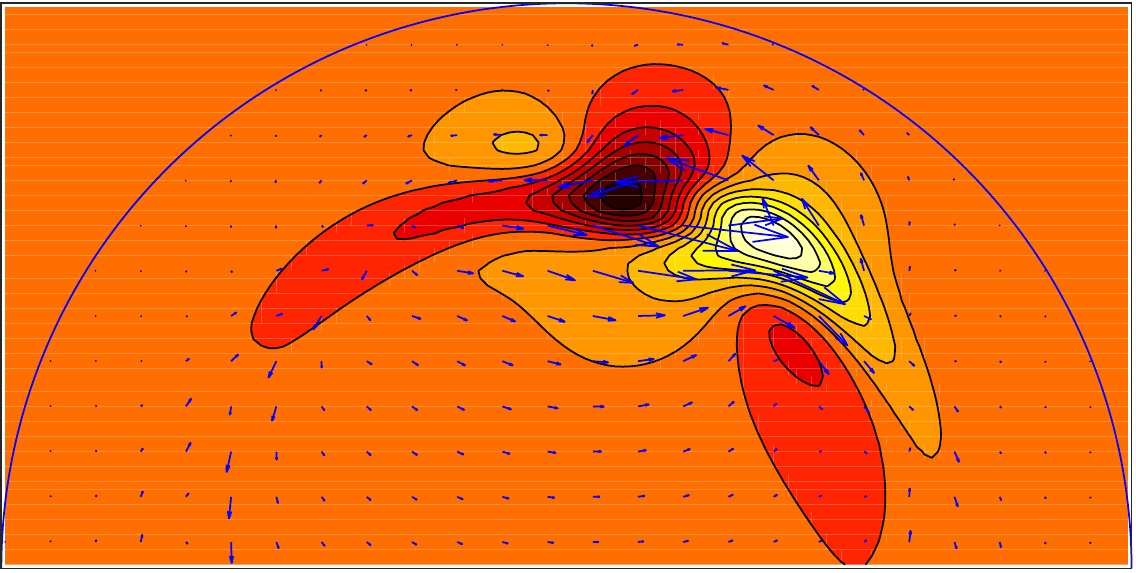}\\
        \includegraphics[width=0.3\textwidth]{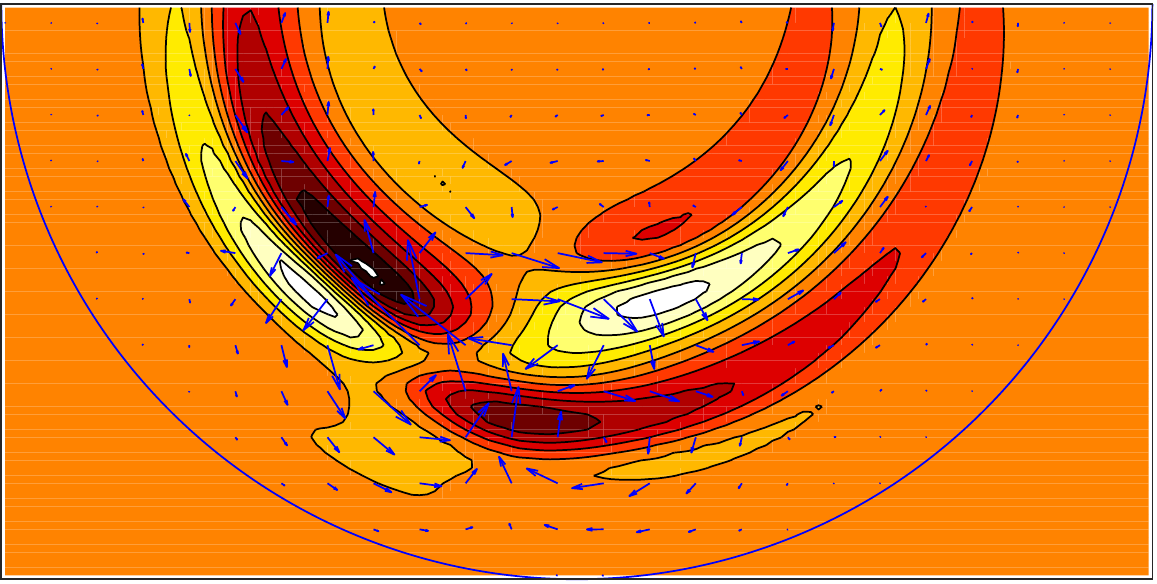}
	\includegraphics[width=0.3\textwidth]{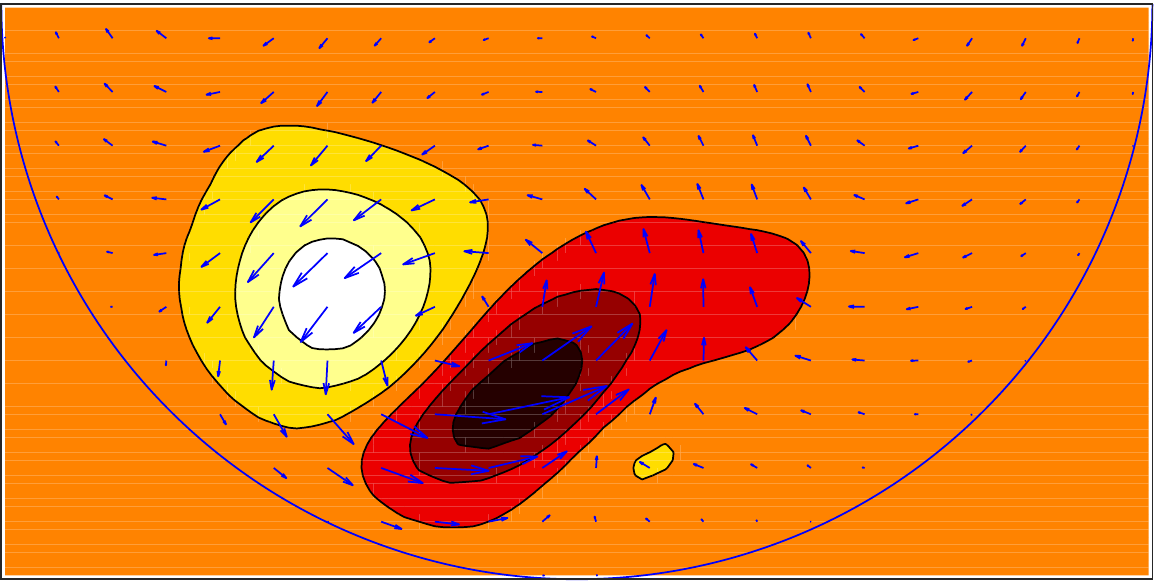}
	\includegraphics[width=0.3\textwidth]{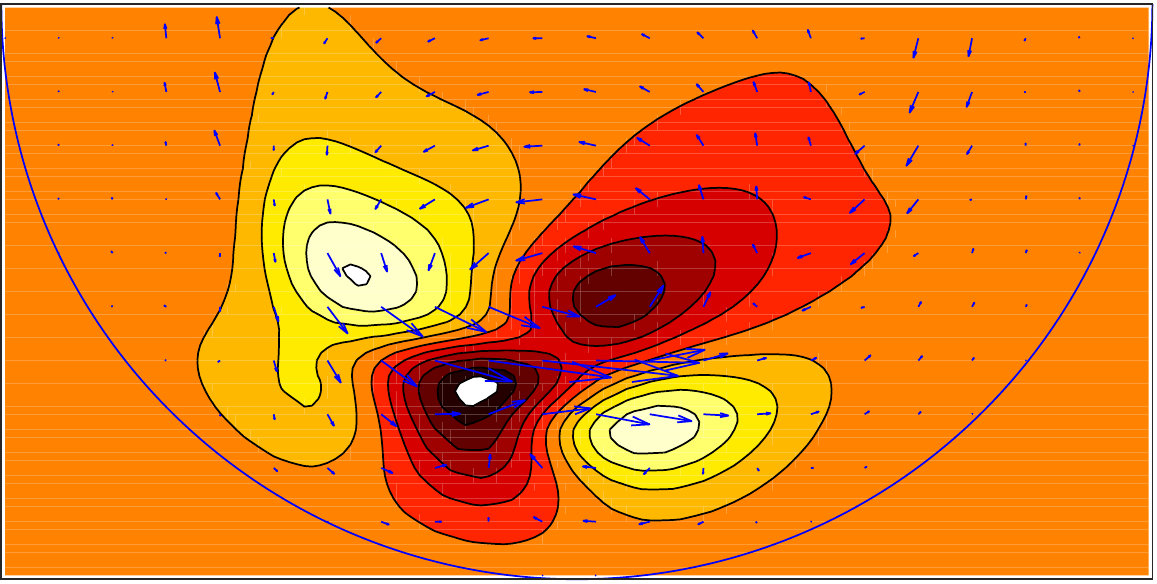}
  \caption{\small Top half: time evolution of the minimal seed for $Re=10000$. From left to right: $t=0$, $t=1.25D/\overline{W}$ (end of the Orr phase) and $t=3.75D/\overline{W}$ (end of the oblique phase). Bottom half: time evolution of the minimal seed (rotated by $\pi$) for $Re=2400$. From left to right: $t=0$, $t=1.1D/\overline{W}$ (end of the Orr phase) and $t=2.5D/\overline{W}$ (end of the oblique phase). The cross sections are taken at the streamwise location of maximum enstrophy. In the left half sections ($t=0$) and right half sections (end of oblique phase), ten contours are used between the extremes of the corresponding streamwise velocity perturbations. The central half sections (end of Orr phase) are scaled with the extremes of the corresponding streamwise velocity perturbations at the end of the oblique phase (right half sections). The cross-sectional velocities (indicated with arrows) are scaled differently for visualisations reasons.}
\label{fig-cshalf}
\end{figure}

As suggested by \citet{kerswell-2018} and explained using a simple example in appendix B of \citet{kerswell-etal-2014}, the minimal seed couples together (via the nonlinear effects) the Orr, oblique-wave and lift-up mechanisms, which occur on different time scales and are uncoupled in the linearised dynamics. By ensuring that the energy of the preceding phase feeds into the following, the minimal seed is thus able to produce a much larger overall growth than any possible in the linearised problem. Our data support this picture and suggest that the oblique-wave process produces a growth of almost $\mathcal{O}(Re)$, which is then further magnified by the lift-up mechanism up to an edge state whose energy scales approximately as $\sim$$Re^{-1}$. From this, it follows that the lift-up mechanism only produces an energy growth of approximately $\mathcal{O}(Re)$, rather than the usually quoted growth factor of $\mathcal{O}(Re^2)$. 
A possible explanation follows from the length scales of the minimal seed becoming finer (and thus the rolls experiencing more dissipation) as the Reynolds number increases. This is evidenced by the cross sections shown in figure \ref{fig-cshalf} of the time evolution of the minimal seed up to the beginning of the lift-up phase for $Re=2400$ and 10000. Rolls advect the mean shear to drive high and low-speed streaks.
The diffusion term for a roll of spanwise wavelength $\ell$ suggests
that such a roll survives a time $\sim Re\,\ell^2$.
For a shear of $\mathcal{O}(1)$, the growth in amplitude of a streak 
is then $\sim Re\,\ell^2$.  The usual argument with 
$\ell=\mathcal{O}(1)$ then implies an energy growth $\sim Re^2$.  Here,
an energy growth $\sim Re$ suggests a length scale $\ell\sim Re^{-1/4}$.

It appears that we do not yet see scaling with wall units,
$\ell^+=Re_\tau \ell$ and $u^+=Re/Re_{\tau}u$, where $Re_{\tau} \equiv u_{\tau} R /\nu$ is the wall Reynolds number ($R$ is the radius of the pipe) and
$u_{\tau}=\sqrt{\nu\p \tilde{w}/ \p r}_{|wall}$ is the friction velocity.
For a localised perturbation ($l^+$ constant) we have $Re_{\tau} \approx \sqrt{2Re}$.
This would suggest a scaling $\ell\sim Re^{-1/2}$. 
In the early stages of the minimal seed dynamics, however, 
assuming that the energy in wall units $E^+$ of a localised perturbation is constant 
leads to a energy scaling $E\sim Re^{-2.5}$.
We observe a scaling not too far off at $E\sim Re^{-2.8}$.  

The energy of the minimal seed at the end of the oblique phase, and 
the localised random initial conditions have similar energies over the range studied.
In this sense, we will regard the minimal seed at the end of the oblique phase
as a reasonable proxy for the transition threshold for random disturbances.

%------------------------------------------------------------------------------------------------------------------------------------------------------------------------
\subsection{Control via a body forcing}
Motivated by the recent experiments performed by Hof and his group, as discussed in \S \ref{intro:control}, we study the effect of adding a localised body force that mimics the presence of a baffle in the core of the flow. 
To study the influence of the baffle on the transition threshold, 
following the results of the previous section,
we use the minimal seed at the end of the oblique phase as an initial condition,
and verify that it measures transition similarly to  
random localised disturbances.

Consider a mesh of stationary point objects in the flow and assume that each point experiences a drag proportional to the total velocity.
The baffle is then approximated by the following forcing
\begin{equation}
\mathbf{F}(r,\theta,z,t)=-A\, \mathcal{B}(z) \, \mathbf{u}_{tot}(r,\theta,z,t)\,,   
\label{forc-drag-tot}
\end{equation}
where $A$ is the (scalar constant) amplitude of the forcing and $\mathcal{B}(z)$ is a (scalar) smoothed step-like function (refer to equation \eqref{smoothing-function}) that introduces a streamwise localisation of the force. The product $A\, \mathcal{B}(z)$ is a measure of the blockage by the fine mesh. The form of the forcing in \eqref{forc-drag-tot} represents a primitive implementation of an immersed boundary method \citep[e.g.][]{peskin-2002,mittal-iaccarino-2005}.
% We introduce a streamwise localisation of the force through a smoothed step-like function $\mathcal{B}(z)$ (refer to equation \eqref{smoothing-function}).
As a first approximation, we assume that the baffle is uniform in the radial direction. The streamwise modulation $\mathcal{B}(z)$ is fixed so that the baffle occupies a fifth of the pipe, 
and the smoothing effect is felt for a tenth of the pipe upstream and downstream of it. The only control parameter is thus $A$. Simulations are fed with the minimal seed at the end of the oblique phase and the random localised disturbance obtained in the unforced cases (refer to figure \ref{fig-statistical-study}), with the energy gradually rescaled until transition is triggered. The time horizon is $T=125(D/\overline{W})$, as in the statistical study presented in section \S \ref{sec:statistical}. For the random localised disturbance we apply a random $z$-shift to the unforced field before feeding it into the simulations with forcing.
For comparison, we have also calculated the effect of the baffle on the transition threshold using the (unforced) minimal seed as initial condition. In this instance, we consider the cases where the initial disturbance is centred in the baffle and half-length away from it.

Our purpose is to investigate whether the flow can be kept laminar in the presence of the baffle and how much net energy can be saved. The presence of the baffle causes a pressure drop downstream, which is measured by $(1+\beta)^A_{lam/turb}=\dissip^A_{lam/turb}/\dissip_{lam}$, where $\dissip^A_{lam/turb}$ is the observed value of the dissipation in the presence of the forcing in either the laminar or turbulent case and $\dissip_{lam}$ is the corresponding laminar value in the unforced case. Hereinafter, the superscript `$A$' indicates the forced case ($A>0$) and the subscripts `$lam$' or `$turb$' refer to the flow being laminar ($E_0<E_c$) or turbulent ($E_0> E_c$) at the current value of $A \ge 0$. We use the turbulent dissipation $\dissip_{turb}$ in the unforced case as a reference value to quantify the effect of the forcing. In the unforced case, $1+\beta\equiv Re_p/Re$ where $Re_p = W_{cl}R/\nu$ is the Reynolds number for fixed pressure ($W_{cl}$ is the centerline velocity of the laminar flow). From the Blasius formula \citep{blasius-1913} for the turbulent friction coefficient $C_f\equiv 16 Re_p/Re^2=0.0791 Re^{-0.25}$, it follows that
\beq
 (1+\beta)_{turb}\equiv\frac{\dissip_{turb}}{\dissip_{lam}}=\frac{Re_p}{Re}=\frac{0.0791}{16} Re^{0.75}\,.
\label{blasius-beta}
\eeq
For the forcing to be beneficial, the dissipation in the presence of the forcing in either the laminar or turbulent case must be lower than the turbulent dissipation in the unforced case, i.e. $\dissip^A_{lam/turb}<\dissip_{turb}$, or, equivalently $(1+\beta)^{A}_{lam/turb}<(1+\beta)_{turb}$. As a measure of how beneficial the forcing is, we define a `laminar' and a `turbulent' drag reduction as:  
\begin{equation}
\mathcal{DR}_{lam/turb}(\%)=\frac{\dissip_{turb} - \dissip_{lam/turb}^{A}}{\dissip_{turb}}=\frac{(1+\beta)_{turb}-(1+\beta)^{A}_{lam/turb}}{(1+\beta)_{turb}}\,.
\label{DR}
\end{equation}

As $A$ is increased, the critical initial energy $E_c$ can be pushed further from the corresponding value in the unforced case, but the pressure drop also increases. For example, in the case of a random localised initial condition, a forcing with $A=0.005$ avoids turbulence being triggered for values of the initial energy where turbulence was first hit in the unforced case (i.e. the initial energies corresponding to the blue crosses in figure \ref{fig-statistical-study}), and a considerable laminar drag reduction is obtained, as shown in table \ref{table-beta}. However, a slight increase of $E_0$ would result in the flow to become turbulent again, with almost no drag reduction being achieved. Hence, with this very low choice of $A$, an almost null raise of $E_c$ is achieved.
\begin{table}
\begin{center}
\begin{tabular}{ c c c c}
 $Re$ & $(1+\beta)_{turb}$ & $(1+\beta)^{A=0.005}_{lam}$ & $\mathcal{DR}_{lam}(\%)$\\
\hline
 2400 & 1.695 & 1.143 &32.5\%\\
 3500 & 2.250 & 1.202 &46.5\% \\ 
 5000 & 2.940 &  1.275 &56.6\%\\
 7000 & 3.783 & 1.354 &64.2\%\\  
 10000 & 4.944 & 1.48 &70\%
\end{tabular}
\caption{Effect of a forcing of amplitude $A = 0.005$ at different Reynolds numbers. The second and fourth columns are calculated using \eqref{blasius-beta} and \eqref{DR}, respectively.}
\label{table-beta}
\end{center}
\end{table}

\begin{figure}
  \centering
%	\includegraphics[width=0.7 \textwidth]{./fig-baffle-Re5000.pdf}\\
%	\hspace{1.3cm}\includegraphics[width=0.7 \textwidth]{./fig-dissip-baffle-Re5000.pdf}
\includegraphics[width=0.8 \textwidth]{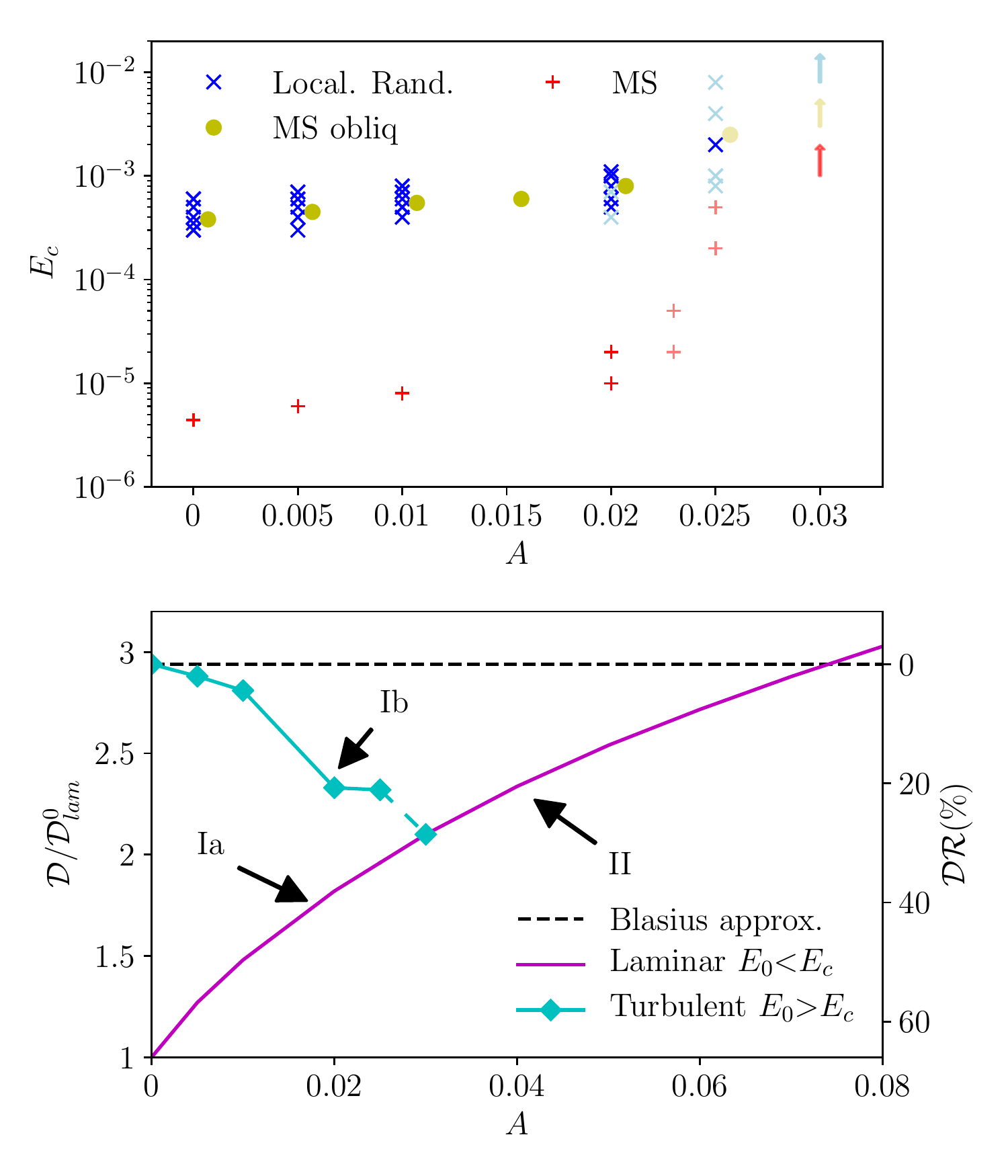}
  \caption{\small Effect of the forcing for different $A$ at $Re=5000$. Top: critical initial energies vs forcing amplitude. The critical initial energies for $A=0$ coincide with the data at $Re=5000$ of figure \ref{fig-statistical-study} for the same types of disturbance. The blue crosses and yellow circles pertain to simulations fed with a random localised initial condition and the minimal seed at the end of the oblique phase, respectively. Note that the yellow circles are slightly shifted for visualisations reasons. In addition, calculations were performed with the minimal seed (with two different shifts applied) as initial disturbance (red plusses).
The dark-coloured symbols indicate cases where the flow remains turbulent as $E_0$ is increased further from the first appearance of turbulence, while the light-coloured symbols denote cases where turbulence is intermittent and characterised by short lifetime. The arrows pointing upwards indicate that $E_c\to\infty$, i.e. a full collapse of turbulence is obtained. Bottom: dissipation and drag reductions vs forcing amplitude. For $A<0.03$ either laminar (Ia: $E_0<E_c$) or turbulent (Ib: $E_0 > E_c$) drag reductions are possible, for $A \ge 0.03$, only laminar drag reduction (II) is achieved as turbulence is suppressed.}
  \label{fig-baffle-Re5000}
\end{figure}

At $Re=5000$ we perform a parametric study on $A$ to find the optimum value which provides the largest $E_c$ at the minimum cost, i.e. with the maximum drag reduction. The results are summarised in figure \ref{fig-baffle-Re5000}. In the top graph the critical energies are shown as a function of $A$ for the initial conditions considered here. The corresponding curves for the turbulent and laminar drag reductions are shown in the bottom graph. Figure \ref{fig-baffle-Re5000} (top) shows that, as $A$ is gradually increased up to $A\approx 0.02$, the critical initial energy increases only slightly, but a considerable turbulent drag reduction is obtained. For example, at $A=0.02$, we obtain $\mathcal{DR}_{turb}=20.7\%$ and $\mathcal{DR}_{lam}=37.8\%$.

\begin{figure}
  \centering
	\subfloat{\includegraphics[width=0.499 \textwidth]{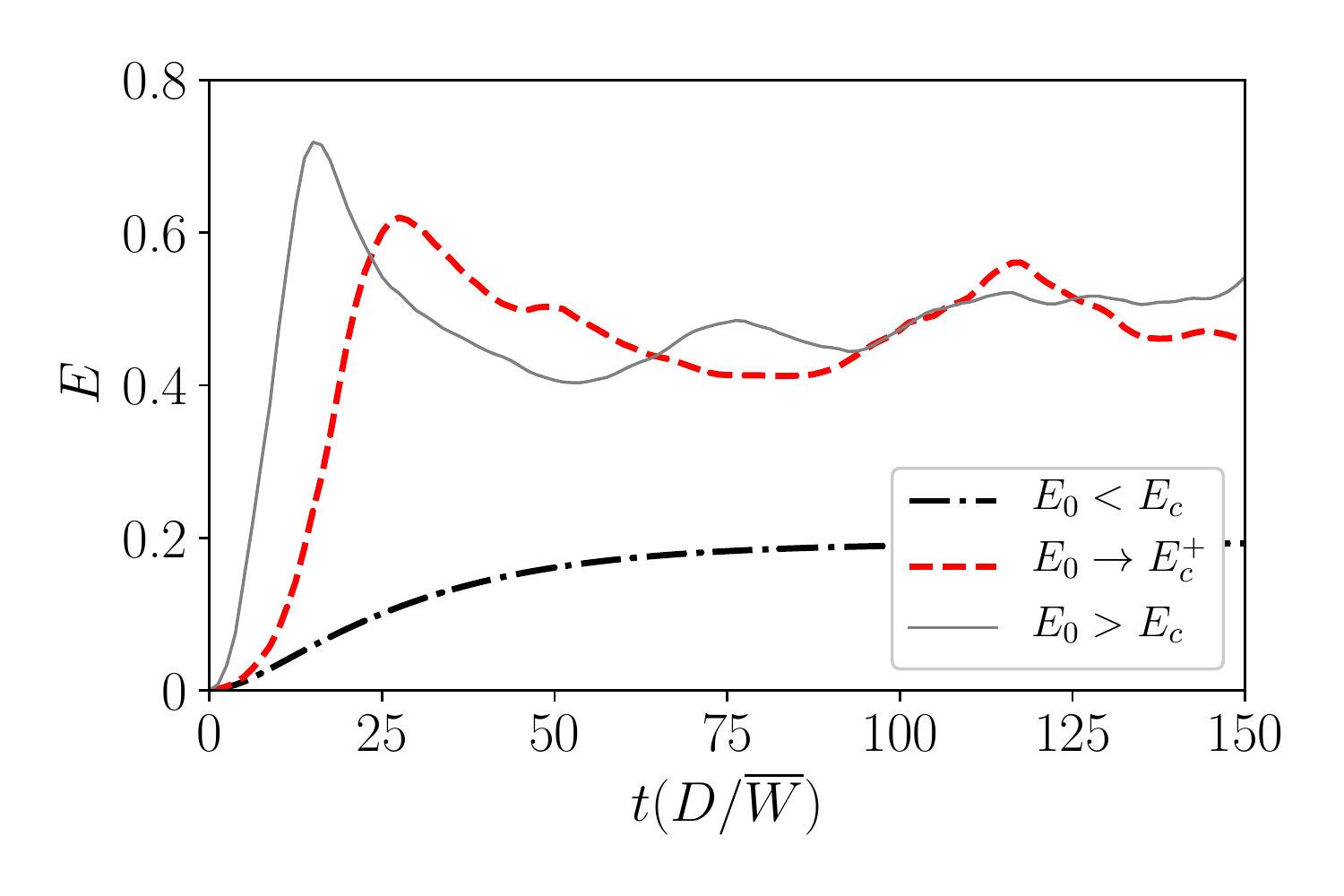}}
	\subfloat{\includegraphics[width=0.499 \textwidth]{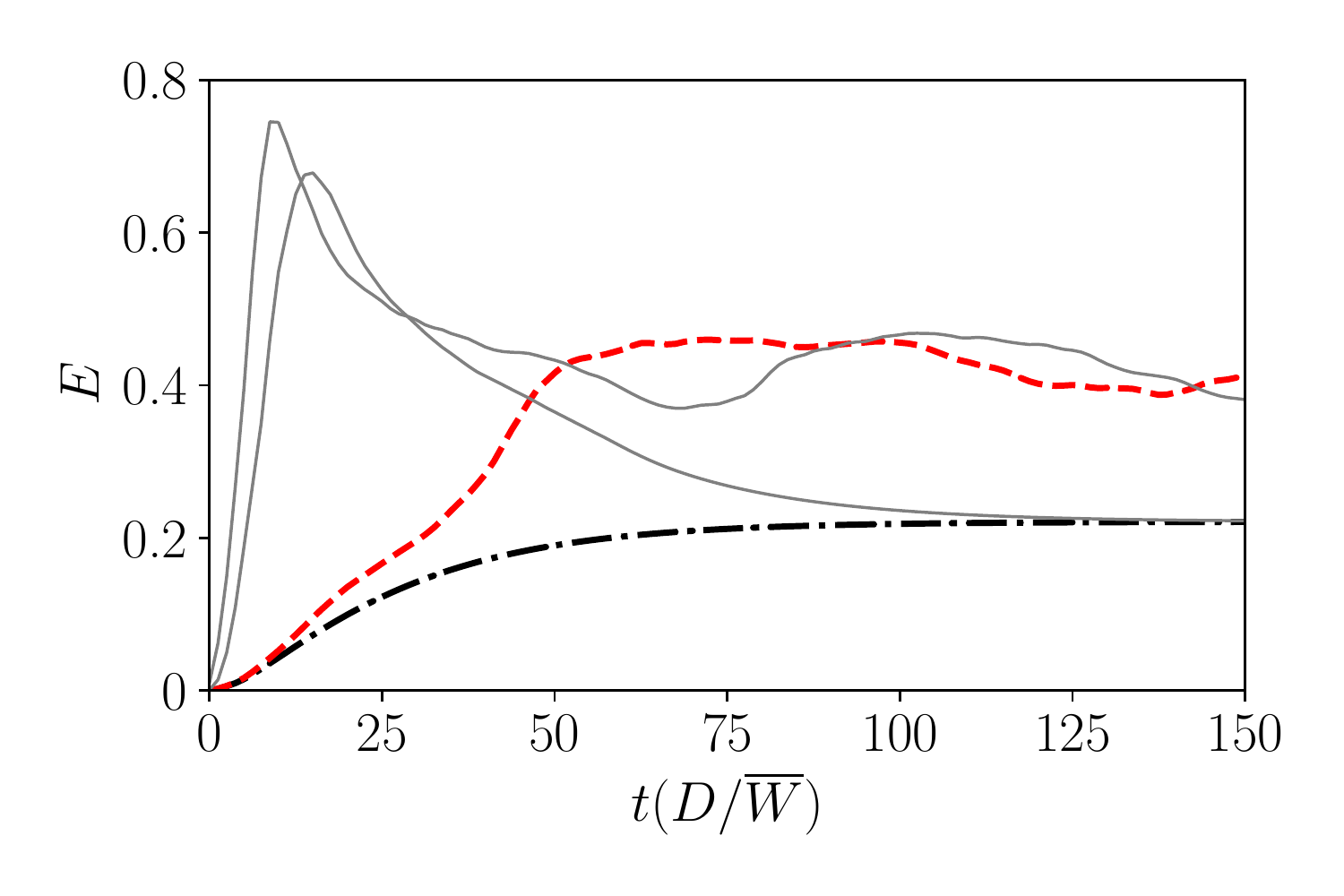}}
  \caption{\small Time series of energy for simulations fed with a random localised disturbance in the presence of a forcing of amplitude $A=0.02$ (left) and $A=0.025$ (right). The left graph shows a case where the flow remains turbulent as $E_0$ is increased further from the first appearance of turbulence (indicated with red dashed line) and turbulence is sustained, while the right graph shows a case where turbulence is intermittent and characterised by short lifetime. In the latter case, for $E_0 > E_c$, the flow is found to either relaminarise or remain turbulent but with a rapid decay towards the end of the observation time window $T=125(D/\overline{W})$. These scenarios are marked with dark and light symbols in figure \ref{fig-baffle-Re5000}, respectively.}
  \label{fig-decay}
\end{figure}

For $A >0.02$, transition starts to become intermittent, i.e. as $E_0$ is increased further from the first appearance of a turbulent episode, the flow either relaminarise or is characterised by short-lifetime turbulence. In these cases, the `critical initial energy' is indicated by light-coloured symbols in figure \ref{fig-baffle-Re5000} (top) to distinguish them from the cases, indicated with dark-coloured symbols, where the flow remains turbulent once turbulence is hit
%EM COMMENTED THE FOLLOWING 
% (verified up to $E_0=\mathcal{O}(10^{-1})$ which corresponds to the order of magnitude of the perturbation energy in the unforced case)
 and turbulence is sustained. 
%EM COMMENTED THE FOLLOWING
%(verified up to $T=250(D/\overline{W})$ which is double the horizon time used in the statistical study).
 An example of these two situations is shown in figure \ref{fig-decay}: in the left graph, the flow remains turbulent for $E_0>E_c$ and turbulence is sustained, while in the right graph some initial disturbances relaminarise and some trigger transition, with the disturbance decaying towards the end of the time window in the latter cases. The scenario where turbulence is short-lived and intermittent is analogous to cases where the Reynolds number is close to the first appearance of turbulence ($Re=1800-2000$), that is, the effect of increasing $A$ is analogous to that of decreasing the Reynolds number. As $A$ is further increased (e.g. at $A=0.025$) this effect becomes more and more pronounced (for example, in the case of the random localised initial condition the `light crosses' become dominant with respect to the `dark crosses')
%EM COMMENTED THE FOLLOWING
% and for all three types of initial conditions the windows of initial energies were the flow relaminarises for $E_0>E_c$ become of greater extent, with only `sporadically' episodes of short-lived turbulence being observed)
 until for $A \ge 0.03$ a full collapse of turbulence is obtained (no turbulence episodes are observed). Therefore, the forcing does significantly modify the basin of attraction of the laminar state by expanding it while making its fractal nature more evident until, first, the chaotic attractor transitions back to a chaotic saddle and finally the laminar state remains the only global attractor. The fact that collapse of the turbulent attractor is possible for a forcing of this form can be confirmed using an energy stability type of analysis (refer to the Appendix): for large enough $A$, the laminar state becomes the global attractor.

At $A=0.03$, where full relaminarisation is first obtained, the laminar drag reduction is still significant (approximately 30$\%$), as shown in the bottom graph of figure \ref{fig-baffle-Re5000}. Therefore we can conclude that this is the optimum choice of forcing amplitude $A_{opt}$. For $A \ge 0.03$ it is not possible to determine $\mathcal{DR}_{turb}$ as turbulence is not observed and thus the curve $\mathcal{DR}_{turb}(A)$ is connected onto $\mathcal{DR}_{lam}(A)$. For $A<0.03$ we can have either laminar or turbulent drag reduction (i.e. the dynamics either sits on curve Ia or Ib in figure \ref{fig-baffle-Re5000}), while for $A \ge 0.03$, we only have laminar drag reduction (the dynamics sits on curve II) due to the relaminarisation (destabilisation of turbulent state). The forcing is found to be beneficial up to $A_c=0.073$, where $\mathcal{DR}_{lam}$ becomes negative, i.e. for $A>A_c$ the cost of the control due to the pressure drop downstream of the baffle becomes greater than the gain due to the relaminarisation. 

Figure \ref{fig-baffle-Re5000} provides further confirmation that the energy of the minimal seed at the end of the oblique phase is a reasonable proxy to measure transition threshold for random localised disturbances. In this sense, the minimal seed at the end of the oblique phase is a useful tool, potentially enabling us to characterise the critical energy using a single simulation in place of a more expensive statistical study.

%%%%%%%%%%%%%%%%%%%%%%%%%%%%%%%%%%%%%%%%%%%%%%%%%%%%%%%%%%%%%%%%%%%%%%%%%%%%%%%%%%%%%%%%%%%%%%%%%%%%%%%%%%%%%%%%%%%%%%%%%%%%%%%%%%%%%%%%%%%%%%%%%%%%%%%%%%%%%%%%%%%%%%%%%%%
\section{Conclusions}
Nonlinear variational methods were used to seek the minimal seed, i.e. the initial perturbation of lowest energy that triggers transition to turbulence. The minimal seed represents the most dangerous disturbance to the basic state and, as a result, is of fundamental interest either from the viewpoint of triggering transition efficiently or, oppositely, in designing flow control strategies. We showed that the structure of the minimal seed is fairly robust to changes in the base flow and to spectral filtering. In the first case, the minimal seed was calculated with a prescribed base flow characterised by a flatter profile in the centre of the pipe as compared to the unforced parabolic profile. In the second case, we projected the initial condition onto a subspace where only a fraction of the streamwise and azimuthal modes were retained. The critical initial energy of the minimal seed was shown to increase with the flatter base profile and with severe spectral filtering (less than 10\% of the modes retained), but the structure of the minimal seed was found to remain largely unchanged in both cases.
%EM COMMENTED THE FOLLOWING
%, thus suggesting that the structure does not have to be `perfectly' formed to be the optimal. 

In order to generate initial conditions that may be considered to model ambient perturbations, we compared the transition behaviour of the minimal seed with that of scaled turbulent snapshots and artificially generated global and localised random disturbances. The random disturbances were obtained by scattering energy randomly over a subset of wavenumbers 
%EM COMMENTED THE FOLLOWING
(the smallest subset for which the critical initial energy of the minimal seed was found to remain unchanged in the previous analysis)
 and calculating the probability of transition as a function of the initial energy in the range of Reynolds numbers from 2400 to 10000. Power-law scalings $E_c=Re^{-\gamma}$ were obtained with $\gamma$ in the range $2-3$ for different forms of disturbances and $\gamma=2.83$ for the minimal seed. The critical initial energy of the minimal seed was found to be approximately two orders of magnitude lower than that of a localised random disturbance, thus suggesting that, despite being robust, the minimal seed is also quite special. However, when we considered the energy of the minimal seed after the initial nonlinear unpacking phase (composed of the Orr and oblique-wave phases), which occurs in a relatively short time scale, the energy gap with the random localised disturbance became negligible. In this sense, the minimal seed at the end of the oblique phase can be regarded as a reasonable proxy for measuring transition thresholds. Energy growth factors of approximately $\mathcal{O}(1)$ and at least $\mathcal{O}(Re)$ for the Orr and oblique-wave mechanisms, respectively, were numerically obtained in the present study for the first time, and the overall picture of the nonlinear growth undergone by the minimal seed to trigger transition discussed. The Orr phase is inviscid and thus the growth produced via this mechanism is independent of the Reynolds number. The oblique-wave process produces a growth in energy of at least $\mathcal{O}(Re)$, which is then seeded to the lift-up mechanism. The disturbance grows by a factor of $\mathcal{O}(Re)$ via the lift-up mechanism, rather than the usually quoted growth factor $\mathcal{O}(Re^2)$ (a reasonable explanation for this is suggested by providing evidence that the structures are getting smaller and smaller as the Reynolds number increases), up to an edge state whose energy is shown to scale as $\mathcal{O}(Re^{-1})$. 

This analysis prepared us with initial conditions for the study of stabilised pipe flows, where a body force was added to the governing equations to mimic the presence of a baffle in the core of the flow, as in the recent experiments by \citet{hof-etal-2010,kuhnen-etal-2018a,kuhnen-etal-2018b}. A parametric study on the effect of the amplitude $A$ of the forcing (corresponding, roughly speaking, to a level of blockage in the pipe due to the baffle) at $Re=5000$ was performed by feeding the simulations with a random localised disturbance and with the minimal seed at the end of the oblique phase found in the unforced case. This confirmed that the minimal seed evolved until the end of the oblique phase is a good proxy for a localised random perturbation, i.e. the critical energy for transition is similar under a variety of forcing situations. An optimum value of the forcing amplitude, $A_{opt}=0.03$, which provides a full collapse of turbulence ($E_c\to \infty$) with a drag reduction of approximately $30\%$, was obtained. Significant drag reductions were found to be possible even in cases where a full collapse of turbulence was not achieved. The forcing was found to be beneficial up to $A_c=0.073$, for values greater than which the cost of the control due to the pressure drop downstream of the baffle exceeded the energy gain. Although it is not possible at this stage to obtain meaningful estimates of $A$ in laboratory experiments \citep[e.g.][]{{kuhnen-etal-2018b}}, due to the artificial forcing used here, this study showed that modifying the core of the flow by inserting an obstacle could be an efficient way of delaying or suppressing turbulence. This method is potentially very attractive as it is passive (no energy input) and very easy to implement. Therefore, we hope that the encouraging results presented here will motivate future developments, such as more realistic modelling of the experimental baffle, to fully exploit the benefits of this control method.
%%%%%%%%%%%%%%%%%%%%%%%%%%%%%%%%%%%
\section{Acknowledgements}
This work was funded by EPSRC grant EP/P000959/1. Fruitful discussions with Bj{\"o}rn Hof are kindly acknowledged. We would like to thank Zijing Ding, Jakob K{\"u}nhen, Chris Pringle, Pierre Ricco and Daniel Wise for insightful comments on a first draft of the manuscript. E.M. would like to acknowledge the assistance of Daniel Wise in producing some of the figures and dealing with technical issues with the computer clusters. This work would have not been possible without the use of the computing facilities of N8 HPC Centre of Excellence, funded by the N8 consortium and EPSRC (Grant EP/K000225/1). The Centre is coordinated by the Universities of Leeds and Manchester.
%%%%%%%%%%%%%%%%%%%%%%%%%%%%%%%%%%%

\appendix
\addcontentsline{toc}{part}{\appendixname}
\noindent
\section*{Appendix. Energy stability analysis of the forced flow}
Consider the Navier-Stokes equations with the forcing given by \eqref{forc-drag-tot}:
\begin{equation}
\frac{\partial \mathbf{u}_{tot}}{ \partial t} + \mathbf{u}_{tot} \cdot \nabla \mathbf{u}_{tot} + \nabla p = \frac{1}{Re}\nabla^2\mathbf{u}_{tot} -A\, \mathcal{B}(z)\,\mathbf{u}_{tot}\,.
\end{equation}
where $\mathbf{u}_{tot}(\mathbf{x},t)= \mathbf{U}(r,z)+\mathbf{\tilde{u}}(\mathbf{x}, t)$, $\mathbf{U}(r,z)$ is some steady basic state (the laminar response to an axisymmetric baffle) and $\mathbf{\tilde{u}}(\mathbf{x}, t)$ is a possibly large perturbation. The latter is governed by:
\begin{equation}
\frac{\partial \mathbf{\tilde{u}}}{ \partial t} + \mathbf{\tilde{u}} \cdot \nabla \mathbf{U} + \mathbf{U} \cdot \nabla \mathbf{\tilde{u}} + \mathbf{\tilde{u}} \cdot \nabla \mathbf{\tilde{u}} + \nabla \tilde{p} = \frac{1}{Re}\nabla^2\mathbf{\tilde{u}} -A\, \mathcal{B}(z)\,\mathbf{\tilde{u}}\,.
\end{equation}
Taking the volume average and simplifying we obtain:
\begin{equation}
\frac{\partial}{\partial t}  \langle \mathbf{\tilde{u}}^2\rangle = 
\langle |\nabla \mathbf{\tilde{u}}|^2\rangle\left \{ \frac{\langle \mathbf{\tilde{u}} \cdot \left[ -A \, \mathcal{B}(z)-\nabla \mathbf{U}\right]\cdot \mathbf{\tilde{u}} \rangle}{\langle |\nabla \mathbf{\tilde{u}}|^2\rangle} -\frac{1}{Re} \right\}\,.
\end{equation}
For the disturbance to decay (i.e. $\partial  \langle \mathbf{\tilde{u}}^2\rangle/ \partial t < 0$), the amplitude of the forcing needs to be sufficiently large so that, for $Re \to \infty$, 
\begin{equation}
\langle \mathbf{\tilde{u}} \cdot \left[ -A \, \mathcal{B}(z)-\nabla \mathbf{U})\right]\cdot \mathbf{\tilde{u}} \rangle < 0 \;\;\forall \, \mathbf{\tilde{u}}\, .
\label{condition}
\end{equation}
%\begin{equation}
%\langle \mathbf{\tilde{u}} \cdot \left[ -A \, \mathcal{B}(z)-\nabla \mathbf{U})\right]\cdot \mathbf{\tilde{u}} \rangle < \frac{\langle |\nabla \mathbf{\tilde{u}}|^2\rangle}{Re} \;\;\forall \, \mathbf{\tilde{u}}\, ,
%\label{condition}
%\end{equation}
%We look for $\frac{\partial}{\partial t}  \langle \mathbf{\tilde{u}}^2\rangle \to 0$ for $Re \to \infty$ so we need $\langle \mathbf{\tilde{u}} \cdot \left[ -A \, \mathcal{B}(z)-\nabla \mathbf{U})\right]\cdot \mathbf{\tilde{u}} \rangle < 0$  $\forall \, \mathbf{\tilde{u}}$
Provided $||\nabla \mathbf{U}|| $ is bounded as $A \to \infty$, we find 
%. Considering the (more conservative) case where $Re \to \infty$ we obtain 
a critical value of the amplitude
\begin{equation*}
A_{crit} = \frac{|| \nabla \mathbf{U}||}{||\mathcal{B}(z)||}\,.
\end{equation*}
Therefore, a forcing with amplitude $A > A_{crit}$ can stabilise any perturbation or, in other words, the steady basic state becomes a global attractor.
%%%%%%%%%%%%%%%%%%%%%%%%%%%%%%%%%%%%
\bibliographystyle{jfm}
\bibliography{./pipes}

\end{document}